\documentclass[reprint,superscriptaddress,showpacs,preprintnumbers,amsmath,amssymb,aps]{revtex4-1}

\usepackage{graphicx,color}
\usepackage{epsfig}
\usepackage{dcolumn}
\usepackage{bm}
\usepackage{CJK}
\usepackage{mathrsfs,amssymb}

\usepackage{booktabs}
\usepackage{multirow}

\newcommand{\te}{\text{e}}
\newcommand{\ti}{\text{i}}

\newcommand{\td}{\text{d}}

\begin{document}

\title{Relativistic description of nuclear matrix elements in neutrinoless double-$\beta$ decay}

\author{L. S. Song}
 \affiliation{State Key Laboratory of Nuclear Physics and Technology, School of Physics, Peking University, Beijing 100871, China}

\author{J. M. Yao}
 \affiliation{Department of Physics, Tohoku University, Sendai 980-8578, Japan}
 \affiliation{School of Physical Science and Technology, Southwest University, Chongqing 400715, China}

\author{P. Ring}
 \affiliation{Physik Department, Technische Universit\"{a}t M\"{u}nchen, D-85748 Garching, Germany}
 \affiliation{State Key Laboratory of Nuclear Physics and Technology, School of Physics, Peking University, Beijing 100871, China}

\author{J. Meng}
 \affiliation{State Key Laboratory of Nuclear Physics and Technology, School of Physics, Peking University, Beijing 100871, China}
 \affiliation{School of Physics and Nuclear Energy Engineering, Beihang University,
              Beijing 100191, China}
 \affiliation{Department of Physics, University of Stellenbosch, Stellenbosch 7602, South Africa}


\begin{abstract}
\begin{description}
	\item[Background]
	Neutrinoless double-$\beta$ ($0\nu\beta\beta$) decay is related to many fundamental concepts in nuclear and particle physics beyond the standard model. Currently there are many experiments searching for this weak process. An accurate knowledge of the nuclear matrix element for the $0\nu\beta\beta$ decay is essential for determining the effective neutrino mass once this process is eventually measured. 
	\item[Purpose]
	We report the first full relativistic description of the $0\nu\beta\beta$ decay matrix element based on a state-of-the-art nuclear structure model.
	\item[Methods]
	We adopt the full relativistic transition operators which are derived with the charge-changing nucleonic currents composed of the vector coupling, axial-vector coupling, pseudoscalar coupling, and weak-magnetism coupling terms. The wave functions for the initial and final nuclei are determined by the multireference covariant density functional theory (MR-CDFT)  based on the point-coupling functional PC-PK1. Correlations beyond the mean field are introduced by configuration mixing of both angular momentum and particle number projected quadrupole deformed mean-field wave functions.
	\item[Results]
	The low-energy spectra and electric quadrupole transitions in ${}^{150}$Nd and its daughter nucleus ${}^{150}$Sm are well reproduced by the MR-CDFT calculations. The $0\nu\beta\beta$ decay matrix elements for both the $0_1^+\rightarrow 0_1^+$ and $0_1^+\rightarrow 0_2^+$ decays of ${}^{150}$Nd are evaluated. The effects of particle number projection, static and dynamic deformations, and the full relativistic structure of the transition operators on the matrix elements are studied in detail. 
	\item[Conclusions]
	The resulting $0\nu\beta\beta$ decay matrix element for the $0_1^+\rightarrow 0_1^+$ transition is $5.60$, which gives the most optimistic prediction for the next generation of experiments searching for the $0\nu\beta\beta$ decay in ${}^{150}$Nd. 
\end{description}
\end{abstract}


\maketitle

\section{Introduction}

Double-$\beta$ ($\beta\beta$) decay is a second-order weak process in which a nucleus decays to the neighboring nucleus by emitting two electrons and, usually, other light particles~\cite{Vergados2012},
\begin{eqnarray}
    (A,Z)\rightarrow (A,Z+ 2)+2e^-+\text{light particles}.
\end{eqnarray}
Owing to the huge $\beta$ decay background, events of this process could, so far, only be recorded in some even-even nuclei, where the $\beta$ decay is energetically forbidden. There are several $\beta\beta$ decay modes including the two-neutrino double-$\beta$ ($2\nu\beta\beta$) decay mode,
\begin{eqnarray}
  (A,Z)\rightarrow(A,Z+2)+2e^-+2\bar\nu_e,
\end{eqnarray}
and the neutrinoless ($0\nu\beta\beta$) decay mode,
\begin{eqnarray}
  (A,Z)\rightarrow(A,Z+2)+2e^-.
\end{eqnarray}

The $2\nu\beta\beta$ mode is allowed in the standard model (SM), while the existence of the $0\nu\beta\beta$ decay would require to go beyond the SM. Evidence for the $0\nu\beta\beta$ decay would be a proof that neutrinos with definite masses are Majorana particles and that neutrino masses have an origin beyond the SM~\cite{Bilenky2003}. This conclusion is independent of the underlying mechanism governing the weak process~\cite{Schechter1982_PhysRevD.25.2951}.

So far, half-lives of the $2\nu\beta\beta$ decay have been measured in 11 isotopes, which are of the order of $10^{18-24}~\mathrm{y}$~\cite{Barabash2010, Gando2012}. However, the $0\nu\beta\beta$ event has never been seen. Only limits of the half-lives can be drawn from current experiments, which are $T_{1/2}^{0\nu}>10^{21-25}~\mathrm{y}$. Searches for the $0\nu\beta\beta$ signals in the $\beta\beta$ candidates are ongoing or proposed in a number of laboratories around the world (see Refs.~\cite{Vergados2012, Avignone2008, Barabash2011_PhysPartNucl} for comprehensive reviews).

Limits of the half-lives $T_{1/2}^{0\nu}$ drawn from experiments provide stringent limits on the parameters associated with the assumed underlying mechanism. Assuming a long-range interaction based on the exchange of a light Majorana neutrino between two weak interaction vertices and restricting the currents to the standard $(V-A)$ form, the part that is proportional to the neutrino mass will be picked out from the neutrino propagator by the same helicity of the coupled leptonic currents~\cite{Pas1999, Vergados2012}. Therefore,
in this case the associated parameter is the effective Majorana neutrino mass. This is called the mass mechanism. Being regarded as the minimal extension of the SM, the mass mechanism is the most popular assumption in current existing theoretical calculations.

Using the mass mechanism, one expects that the $0\nu\beta\beta$ observation, combined with the results of neutrino oscillation experiments, will allow to obtain important information about the character of the neutrino mass spectrum, about the minimal neutrino mass $m_1$ and about the Majorana Charge-Parity violating phase~\cite{Bilenky1987, Bilenky2003}. To extract the neutrino mass, the inverse half-life can be factorized as
\begin{eqnarray}
    \left[T_{1/2}^{0\nu}\right]^{-1}=G_{0\nu} g_A^4(0)\left|\frac{\langle m_\nu\rangle}{m_e}\right|^2{\left|M^{0\nu}(0_I^+\rightarrow 0_F^+)\right|^2},
\end{eqnarray}
where the axial-vector coupling constant $g_A(0)$ and the electron mass $m_e$ are constants, and the kinematic phase-space factor $G_{0\nu}$ can be determined precisely~\cite{Kotila2012}. Therefore, the accurate knowledge of the nuclear matrix element (NME) $M^{0\nu}$ plays a crucial role for extracting the effective neutrino mass $\langle m_\nu\rangle$ from the measurement of the decay rate.

The calculation of the NME requires two main ingredients. One is the decay operator, which reflects the mechanism governing the decay process. The other is the wave functions of the initial and final states. They are provided by theoretical nuclear models and carry the nuclear structural information. Methods used in the literature to calculate the wave functions include the quasiparticle random phase approximation (QRPA)~\cite{Simkovic1999,Kortelainen2007_PhysRevC.75,Kortelainen2007_PhysRevC.76,Simkovic2008, Fang2010,Fang2011,Mustonen2013}, the interacting shell model (ISM)~\cite{Caurier2008_PRL,Menendez2009,Neacsu2012}, the interacting boson model (IBM)~\cite{Barea2009,Barea2013}, the projected Hartree-Fock-Bogoliubov (PHFB)~\cite{Chaturvedi2008, Rath2010, Rath2013}, and the nonrelativistic energy density functional (NREDF) theories~\cite{Rodriguez2010, Rodriguez2011, Rodriguez2013, Vaquero2013}. In the PHFB, the beyond-mean-field correlation connected with the restoration of broken rotational symmetry is taken into account. In the NREDF, additional correlations connected with particle number projection, as well as fluctuations in quadrupole shapes~\cite{Rodriguez2010} and pairing gaps~\cite{Vaquero2013}, are included. Therefore, this method is also referred to as the multireference density functional theory. All these methods used so far are based on nonrelativistic quantum mechanics. The nonrelativistic reduced transition operators are therefore adopted in the calculations of the NMEs for the neutrinoless double-$\beta$ decay.

In the past decades, covariant density functional theory (CDFT) has been proven to be a very powerful tool in nuclear physics. On the mean-field level, the single-reference CDFT, or the relativistic mean-field (RMF) theory, provides a good description of the static ground-state properties for finite nuclei~\cite{Serot1986, Reinhard1989, Ring1996, Vretenar2005, Meng2006}. The relativistic version of energy density functional (REDF) takes into account Lorentz invariance, which puts stringent restrictions on the number of parameters. The spin-orbit potential is included naturally and uniquely, as well as the time-odd components of the nuclear mean field. With the merits inherited, this method has also been generalized beyond the static mean-field level by the RPA~\cite{Ring2001, Liang2008} and QRPA~\cite{Paar2007, Paar2009, Niu2009, Niu2013} or by the multireference CDFT (MR-CDFT) method~\cite{Niksic2006_PhysRevC.73, Niksic2006_PhysRevC.74, Yao2008, Yao2009, Yao2010, Yao2011,Yao2014}, so that it could be applied for the description of the excited states, electromagnetic properties, and the weak transitions including the single- and double-$\beta$ decay.

Relativistic QRPA calculations based on the CDFT have been carried out for the NMEs of the $2\nu\beta\beta$ decay~\cite{Conti2012}, where the transition operator has the same form as that used in the nonrelativistic studies. However, research in the $0\nu\beta\beta$ mode has still to be done. The purpose of this work is to close this gap and to give a relativistic description for the NMEs of the $0\nu\beta\beta$ decay within the framework of MR-CDFT. First, MR-CDFT is able to give a unified description of all the $0\nu\beta\beta$ candidates including heavy deformed nuclei. Furthermore, reliable wave functions can be provided, with the restoration of symmetries by angular momentum projection (AMP) and particle number projection (PNP), as well as the inclusion of configuration mixing by the generator coordinate method (GCM). In addition, because the wave functions are Dirac spinors, the transition operator derived from the Feynman diagram of weak interaction, which is a $4\times 4$ matrix, can be directly sandwiched between the initial and final states without further reduction. Therefore, this investigation also provides a way of testing the validity of the nonrelativistic reduction for the decay operator adopted in the nonrelativistic studies.

As the first attempt we investigate the $0\nu\beta\beta$ decay of ${}^{150}$Nd, which is one of the most promising candidates for the $0\nu\beta\beta$ decay experiments. It has the second highest endpoint energy ($Q_{\beta\beta}=3.37~\mathrm{MeV}$) and the largest phase-space factor $G_{0\nu}$ for the decay~\cite{Kotila2012}. It does not seem feasible that this heavy deformed nucleus can be treated in the near future by a reliable shell-model calculation. However, research has been done with other methods so that comparisons can be made. In particular, detailed discussion can be found for ${}^{150}$Nd and the daughter nucleus ${}^{150}$Sm in Ref.~\cite{Rodriguez2010}, including the results for the spectra of low-lying excited states, the $E2$ transition probabilities, the collective wave functions, and the NMEs between them. We investigate the same nuclei to have a direct comparison of the results from two different state-of-the-art energy density functional (EDF) methods, one of them nonrelativistic and another relativistic. Previous research has shown that the nuclear deformation is responsible for the suppression of the transition matrix element for ${}^{150}$Nd. Therefore, we pay particular attention to the effects of deformation and the corresponding shape fluctuations. Moreover, ${}^{150}$Nd is one of the two isotopes where the transition to the first $0^+$ excited states of their daughter nuclei have been recorded in the $2\nu\beta\beta$ decay experiments~\cite{Barabash2010}. Therefore, from the experimental point of view, it is interesting to evaluate also the $0_1^+\rightarrow 0_2^+$ transition in addition to the ground-state to ground-state transition.

There have been numerous discussions about the uncertainties in the calculated NMEs related to the closure approximation, the inclusion of the high-order currents and the tensorial part induced by the high-order currents, the treatment of the finite nucleon size correction as well as the short-range correlation, and the use of different renormalized values for the axial-vector coupling constant $g_A(0)$, for instance, in Refs.~\cite{Simkovic1999, Rodin2006, Rodin2007, Kortelainen2007_PhysRevC.76, Rath2010, Rath2013, Simkovic2011}. Because it is not our prior task in this paper to estimate these uncertainties, we just clarify here a few things about our calculations. (1) The matrix elements are calculated in the closure approximation. (2) The high-order currents are fully incorporated and the tensorial part is included automatically in the relativistic formalism. (3) The finite nucleon size correction is taken care of by the momentum-transfer-dependent form factors. (4) According to a recent study~\cite{Simkovic2009_PhysRevC}, realistic values of short range correlation have only a small effect ($<7\%$) on the matrix elements; thus, we omit the contribution of short-range correlation presently. (5) {Investigations~\cite{Menendez2011_PRL, Engel2014} show that the chiral two-body hadronic currents provide important contributions to the quenching of Gamow-Teller transitions. A momentum-transfer dependence for this quenching effect is predicted. Therefore, it is more reasonable to include the contributions of two-body currents approximately (on the one-body level) by introducing an effective $g_A^\text{eff}(\bm q^2)$ than introducing a renormalized constant $g_A^\text{eff}(0)$. Because the study on the effect of chiral two-body currents is far beyond the scope of this paper, the coupling constant is set to $g_A(0)=1.254$ (not to some renormalized values) in the following discussion.}

This paper is organized in the following way. In Sec.~\ref{theory}, the derivation of the $0\nu\beta\beta$ decay operator in the mass mechanism, the formalism of the MR-CDFT, and the expressions for the $0\nu\beta\beta$ decay matrix elements in MR-CDFT are briefly introduced. Section~\ref{numerical} is devoted to the numerical details. In Sec.~\ref{result} we present the results for the nuclear structure properties and the NMEs of the $0\nu\beta\beta$ decay. Last, the investigations are summarized in Sec.~\ref{summary}.

\section{Theoretical Framework}\label{theory}
\subsection{Decay operator}
Derivations of the $0\nu\beta\beta$ decay operator can be found in many papers, such as Refs.~\cite{Simkovic1999, Simkovic2008, Avignone2008}. However, the authors end up with the nonrelativistic reduced operator. Therefore, to have a consistent relativistic description, it becomes necessary to repeat the crucial steps of the derivation and to show the form of the relativistic operator involved in our calculations.

The starting point is the semileptonic charged-current weak Hamiltonian~\cite{Walecka1975},
\begin{eqnarray}
    \mathcal{H}_\text{weak}(x)=\frac{G_F\cos \theta_C}{\sqrt{2}} j^\mu (x)\mathcal J_{\mu}^\dagger(x)+\text{H.c.},
\end{eqnarray}
where $G_F$ is the Fermi constant, $\theta_C$ is the Cabbibo angle, and the standard leptonic current adopts ($V-A$) form:
\begin{eqnarray}
     j^\mu (x)= \bar e (x)\gamma^\mu (1-\gamma_5)\nu_e(x).
\end{eqnarray}
The hadronic current is expressed in terms of nucleon field $\psi$,
\begin{eqnarray}\label{Ncurrent}
  \mathcal J_{\mu}^\dagger (x)&=&\bar \psi (x)\left[g_V(q^2)\gamma_\mu+\ti g_M(q^2)\frac{\sigma_{\mu\nu}}{2m_p}q^\nu\right.\notag\\
  &-&\left.g_A(q^2)\gamma_\mu \gamma_5-g_P (q^2)q_\mu \gamma_5\right]\tau_-\psi(x),
\end{eqnarray}
where
$m_p$ is the nucleon mass, $q^\mu$ is the momentum transferred from leptons to hadrons, $\tau_-$ is the isospin lowing operator, and $\sigma_{\mu\nu}=\frac{\ti}{2}\left[\gamma_\mu,\gamma_\nu\right]$. The form factors $g_V(q^2)$, $g_A(q^2)$, $g_M(q^2),$ and $g_P(q^2)$, in which the effects of the finite nucleon size are incorporated, represent respectively, in the zero-momentum-transfer limit, the vector, axial-vector, weak-magnetism, and induced pseudoscalar coupling constants. We adopt here the same expressions for the form factors as in Ref.~\cite{Simkovic1999}.

By using the long-wave approximation for the outgoing electrons and neglecting the small energy transfer between nucleons, the NME $M^{0\nu}$ of the $0\nu\beta\beta$ decay can be obtained after a few steps~\cite{Bilenky1987},
\begin{eqnarray}\label{M0nu}
  M^{0\nu}(0_I^+\rightarrow 0_F^+)\equiv \langle 0_F^+|\hat{\mathcal O}^{0\nu}|0_I^+\rangle,
\end{eqnarray}
where $|0_{I/F}^+\rangle$ is the wave function of the initial ($I$)/final ($F$) state, and the decay operator reads
\begin{eqnarray}
\label{operator0}
  \hat{\mathcal O}^{0\nu}&=&\frac{4\pi R}{g_A^2(0)} \iint d^3 x_1d^3 x_2 \int\frac{d^3 q}{(2\pi)^3}\frac{\te ^{\ti \bm q\cdot(\bm x_1-\bm x_2)}}{q}\notag\\
  &\times&\sum_m\frac{\mathcal {J}_{\mu}^\dagger(\bm x_1)|m\rangle\langle m|\mathcal {J}^{\mu\dagger}(\bm x_2)}{q+E_m-(E_I+E_F)/2},
\end{eqnarray}
where $R=r_0A^{1/3}$, with $r_0=1.2~\mathrm{fm}$ introduced to make the NME dimensionless. The summation runs over all the possible states $|m\rangle$ of the intermediate nucleus, and $E_m$ is the corresponding energy of each state.

Replacing the state-dependent energy with an average one: $E_m\rightarrow \bar E$, the intermediate states can be eliminated by making use of the relation $\sum_m |m\rangle\langle m|=1$. Then the operator becomes
\begin{eqnarray}
\label{operator1}
  \frac{4\pi R}{g_A^2(0)} \iint d^3 x_1d^3 x_2 \int\frac{d^3 q}{(2\pi)^3}\frac{\te ^{\ti \bm q\cdot(\bm x_1-\bm x_2)}}{q}\frac{\mathcal {J}_{\mu}^\dagger(\bm x_1)\mathcal {J}^{\mu\dagger}(\bm x_2)}{q+E_d},\notag\\
\end{eqnarray}
where $E_d\equiv \bar E-(E_I+E_F)/2$, is the average excitation energy. There are claims that this closure approximation is reliable in the calculation of $M^{0\nu}$, because different values of the energy parameter $E_d$ within a certain range will not lead to dramatic changes of $M^{0\nu}$~\cite{Rodin2006, Rodin2007, Simkovic2011, Rath2013}. The sensitivity of the matrix elements to the changes of $E_d$ is discussed further later.

Considering the four terms in Eq. (\ref{Ncurrent}), the operator can be decomposed into the vector coupling (VV), axial-vector coupling (AA), axial-vector and pseudoscalar coupling (AP), pseudoscalar coupling (PP), and weak-magnetism coupling (MM) channels, as
\begin{eqnarray}\label{O0nu}
    \hat{\mathcal O}^{0\nu}=\sum_{i}\hat{\mathcal O}^{0\nu}_{i},\quad \left(i=VV,AA,AP,PP,MM\right)
\end{eqnarray}
with each component being
\begin{eqnarray}
\label{operator2}
  \hat{\mathcal O}^{0\nu}_i=\frac{4\pi R}{g_A^2(0)}\iint d^3 x_1d^3 x_2 \int\frac{d^3 q}{(2\pi)^3}\frac{\te^{\ti \bm q\cdot(\bm x_1-\bm x_2)}}{q(q+E_d)}
  \left[\mathcal {J}_{\mu}^\dagger\mathcal {J}^{\mu\dagger}\right]_i,\notag\\
\end{eqnarray}
and the ``two-current'' operators $\left[\mathcal {J}^\dagger_\mu\mathcal {J}^{\mu\dagger}\right]_{i}$ being
\begin{subequations}\label{twocurrentR}
\begin{eqnarray}
\label{VV}
&&g_V^2(\bm q^2)\left(\bar\psi\gamma_\mu\tau_-\psi\right)^{(1)}\left(\bar\psi\gamma^\mu\tau_-\psi\right)^{(2)},\\
\label{AA}
&&g_A^2(\bm q^2)\left(\bar\psi\gamma_\mu\gamma_5\tau_-\psi\right)^{(1)}\left(\bar\psi\gamma^\mu\gamma_5\tau_-\psi\right)^{(2)},\\
\label{AP}
&&2g_A(\bm q^2)g_P(\bm q^2)\left(\bar\psi\bm \gamma\gamma_5\tau_-\psi\right)^{(1)}\left(\bar\psi \bm q\gamma_5\tau_-\psi\right)^{(2)},\\
\label{PP}
&&g_P^2(\bm q^2)\left(\bar\psi \bm q\gamma_5\tau_-\psi\right)^{(1)}\left(\bar\psi \bm q\gamma_5\tau_-\psi\right)^{(2)},\\
\label{MM}
&&g_M^2(\bm q^2)\left(\bar\psi\frac{\sigma_{\mu i}}{2m_p}q^i\tau_-\psi\right)^{(1)}\left(\bar\psi\frac{\sigma^{\mu j}}{2m_p}q_j\tau_-\psi\right)^{(2)}.~~~~~~~~
\end{eqnarray}
\end{subequations}

\subsection{Nuclear wave function}
This work is based on the MR-CDFT, discussed in detail in Ref.~\cite{Yao2010}, taking into account the symmetry restoration by the projection method and the configuration mixing by the GCM.  Therefore, the wave functions for the initial and final nuclei in Eq.~(\ref{M0nu}) are derived by the MR-CDFT calculations. The trial projected GCM wave function $|JMNZ;\alpha\rangle$ reads~\cite{Yao2014}
\begin{eqnarray}\label{trial}
  |JMNZ;\alpha\rangle= \sum_{q,K}f_\alpha^{JK}(q)\hat P_{MK}^J\hat P^N\hat P^Z|q\rangle,
\end{eqnarray}
where $\alpha=1,2,\ldots$ distinguishes different eigenstates of the collective Hamiltonian for given angular momentum $J$, and $|q\rangle$ denotes a set of RMF+BCS states with different quadrupole deformations $q\equiv(\beta,\gamma)$.
The particle number projectors $\hat P^{N_\tau}$ have the form
\begin{eqnarray}
  \hat P^{N_\tau}=\frac{1}{2\pi}\int_0^{2\pi}d \varphi_\tau \te^{\ti\varphi_\tau(\hat N_\tau-N_\tau)}\quad(\tau=n,p),
\end{eqnarray}
and the operators $\hat P_{MK}^J$ for three-dimensional AMP are
\begin{eqnarray}
  \hat P_{MK}^J=\frac{2J+1}{8\pi^2}\int d\Omega D_{MK}^{J*}(\Omega)\hat R(\Omega),
\end{eqnarray}
where $\Omega$ represents the Euler angles $(\phi, \theta, \psi)$, and the measure is $d \Omega=d \phi \sin\theta d\theta d\psi$. $D_{MK}^J(\Omega)$ is the Wigner $D$ function. The rotational operator is chosen in the notation of Edmonds~\cite{Edmonds1957}: $\hat R(\Omega)=\te^{\ti\phi\hat J_z}\te^{\ti\theta\hat J_y}\te^{\ti\psi\hat J_z}$.

The weight functions $f_\alpha^{JK}(q)$ in the wave function of Eq.~(\ref{trial}) are determined by requiring that the expectation value of the Hamiltonian is stationary with respect to an arbitrary variation $\delta f_\alpha^{JK}(q)$, which leads to the Hill-Wheeler-Griffin equation~\cite{Griffin1957},
\begin{eqnarray}
  \sum_{q',K'}\left[\mathscr H_{KK'}^J(q,q')-E_\alpha^J \mathscr N_{KK'}^J(q,q')\right]f_\alpha^{JK'}(q')=0,~~~
\end{eqnarray}
where the kernel function contains a Hamiltonian kernel $\mathscr H_{KK'}^J(q,q')$ and a norm kernel $\mathscr N_{KK'}^J(q,q')$~\cite{Yao2010}.

Solving the above equation as in Ref.~\cite{Yao2010},
one can determine both the energies $E_\alpha^J$ and the amplitudes $f_\alpha^{JK}(q)$,
\begin{eqnarray}
  f_\alpha^{JK}(q)\equiv f_\alpha^J(i)=\sum_k\frac{g_k^{J\alpha}}{\sqrt{n_k^J}}u_k^J(i),
\end{eqnarray}
where the index $i$ has a one-to-one correspondence with the mesh point $(K, q)$ in the $K\bigotimes q$ space and $n_k^J$ and $u_k^J(i)$ are the eigenvalues and the corresponding eigenstates of the norm $\mathscr N^J(i,i')$. $E_\alpha^J$ and $g_k^{J\alpha}$ are the eigenvalues and the corresponding eigenvectors, respectively, of the Hamiltonian constructed with the ``natural states''~\cite{Ring1980} with $n_k^J\neq 0$:
\begin{eqnarray}
	H_{kl}^J=\sum_{ii'}\frac{u_k^{J*}(i)}{\sqrt{n_k^J}}\mathscr H^J(i,i') \frac{u_l^{J*}(i')}{\sqrt{n_l^J}}.
\end{eqnarray}
The collective wave functions $g_\alpha^J(i)$ are constructed as
\begin{eqnarray}
\label{gfunction}
  g_\alpha^J(i)=\sum_k g_k^{J\alpha} u_k^J(i),
\end{eqnarray}
where $g_\alpha^J(i)$ are normalized as $\sum_{i} g_\alpha^{J*}(i) g_{\alpha'}^J(i)=\delta_{\alpha\alpha'}$ and, therefore, $|g_\alpha^J(i)|^2$ can be interpreted as a probability amplitude. More details about the calculations of observables within this framework can be found in Ref.~\cite{Yao2010}.

\subsection{Evaluation of NME}
In the following investigation we concentrate on the wave functions with axial symmetry, with one collective coordinate $q=\beta$, and we restrict ourselves to states with the quantum numbers $J^\pi=0^+$. With the GCM wave functions the NME in Eq.~(\ref{M0nu}) can be expressed as
\begin{equation}
\label{M}
M^{0\nu}=\sum_{\beta_I, \beta_F}f^\ast_{0_F^+}(\beta_F)f_{0_I^+}(\beta_I) M^{0\nu}(\beta_I,\beta_F)
\end{equation}
with the projected NMEs at different deformations:
\begin{equation}
\label{MIF}
M^{0\nu}(\beta_I,\beta_F)=\langle \beta_F|\hat{\mathcal O}^{0\nu}\hat{P}^{J=0}\hat P^{N_I}\hat P^{Z_I}|\beta_I\rangle.
\end{equation}
In these matrix elements we keep explicitly the projection operators on one side of the operator only (single projection), because it is equivalent to the double projection on both sides. To prove this we consider for the sake of simplicity only the projection onto good proton number. In this case the wave function $\hat{P}^{Z}| \beta_I\rangle$ contains only components with proton number $Z$. The operator $\hat{\mathcal O}^{0\nu}$ creates two protons and therefore the wave function $\hat{\mathcal O}^{0\nu}\hat{P}^{Z}| \beta_I\rangle$ has only components with proton number $Z+2$. Applying $\hat{P}^{Z+2}$ onto this function is equivalent with the unity, i.e.,
\begin{equation}
\label{singlePNP}
\langle \beta_F|\hat{P}^{Z+2}\hat{\mathcal O}^{0\nu}\hat{P}^{Z}| \beta_I\rangle=\langle \beta_F|\hat{\mathcal O}^{0\nu}\hat{P}^{Z}| \beta_I\rangle.
\end{equation}
The NME $M^{0\nu}$ in Eq. (\ref{M}) can be regarded as a weighted summation over the matrix elements with different initial and final deformations.
This summation leads, therefore, to configuration mixing in the nuclear wave functions.

The wave function $\hat{P}^{J=0}\hat{P}^{N}\hat{P}^{Z}|\beta\rangle$ in Eq.~(\ref{MIF}) is not normalized. For later convenience and to compare with PHFB calculations~\cite{Rath2010, Rath2013}, we also introduce a single-configuration transition matrix element $\tilde{M}^{0\nu}(\beta_I, \beta_F)$ between the normalized initial and normalized final states with definite deformations $\beta_I$ and $\beta_F$,
\begin{equation}\label{MIFnorm}
\tilde{M}^{0\nu}(\beta_I, \beta_F) = {\mathcal N}_F{\mathcal N}_I\,\langle \beta_F|\hat{\mathcal O}^{0\nu}\hat P^{J=0}\hat P^{N_I}\hat P^{Z_I}| \beta_I\rangle,
\end{equation}
with ${\mathcal N}^{-2}_a= \langle \beta_a|\hat  P^{J=0}\hat P^{N_a}\hat P^{Z_a}| \beta_a\rangle$ for $a=I,F$.
Note that this single-configuration matrix element is normalized at each configuration $(\beta_I, \beta_F)$ with the norm of the two projected states. This quantity gives the results of the PHFB method for the NME. It shows the influence of the nuclear deformations on the strength of the $0\nu\beta\beta$ decay, but it does not take into account fluctuations in deformation space, which are very important in transitional nuclei.

Writing the projection operators explicitly and using the second-quantized form of $\hat{\mathcal O}^{0\nu}$, the matrix element in Eq.~(\ref{MIF}) becomes
\begin{eqnarray}\label{me}
&&M^{0\nu}(\beta_I,\beta_F)=\sum_{abcd}\langle ab|\hat{O}|cd\rangle\notag~~~~~~~~~~~~~~~~~~~~~\\
&&~~~~~~~~~~~\times~\int\limits_0^\pi \frac{\sin\theta d\theta}{2}\int\limits_0^{2\pi}\frac{d\varphi_{n}}{2\pi}\,\te^{-\ti\varphi_{n}N_I}
  \int\limits_0^{2\pi}\frac{d\varphi_{p}}{2\pi}\,\te^{-\ti\varphi_{p}Z_I}\notag\\
&&~~~~~~~~~~~\times~\langle \beta_F|c_a^{(\pi)\dagger}c_b^{(\pi)\dagger}c_d^{(\nu)}c_c^{(\nu)}|\tilde{\beta_I}\rangle,
\end{eqnarray}
where
$c_d^{(\nu)},c_{c}^{(\nu)}$ are neutron annihilation and $c_a^{(\pi)\dagger},c_b^{(\pi)\dagger}$ are proton creation operators. The indices $c,d$ run over a complete set of single neutron states and $a,b$ over a complete set of single proton states. The shorthand notation $|\tilde{\beta_I}\rangle$ stands for
  \begin{eqnarray}
    |\tilde{\beta_I}\rangle\equiv\te^{\ti \theta\hat{J}_y}\te^{\ti \varphi_{n}\hat N}\te^{\ti\varphi_{p}\hat Z}|\beta_I\rangle.
  \end{eqnarray}

The crucial part that contains the nuclear structural information in Eq.~(\ref{me}) is the two-body transition density,
$ \langle \beta_F| c_a^{(\pi)\dagger}c_b^{(\pi)\dagger}c_d^{(\nu)}c_c^{(\nu)}|\tilde{\beta_I}\rangle $.
Provided that the states $|\beta_F\rangle$ and $|\tilde{\beta_I}\rangle$ are not orthogonal, one can use the extended Wick's theorem of Refs.~\cite{Onishi1966,Balian1969} and
express the two-body transition density as a product of a norm overlap and two one-body transition pairing tensors as
\begin{eqnarray}
  &&\langle \beta_F| c_a^{(\pi)\dagger}c_b^{(\pi)\dagger}c_d^{(\nu)}c_c^{(\nu)}|
  \tilde{\beta_I}\rangle=n(\theta,\varphi_{n},\varphi_{p};\beta_I,\beta_F)~~~~~~\notag\\
  &&\times~\kappa^{01*(\pi)}_{ab}(\theta,\varphi_{p};\beta_I,\beta_F)
  \times\kappa^{10(\nu)}_{cd}(\theta,\varphi_{n};\beta_I,\beta_F).
\end{eqnarray}
The norm overlap is given by
\begin{eqnarray}
   n(\theta,\varphi_{n},\varphi_{p};\beta_I,\beta_F)\equiv \langle \beta_F|
\tilde{\beta_I}\rangle,
\end{eqnarray}
and the transition pairing tensor matrices are
\begin{subequations}
\begin{eqnarray}
    \kappa^{01*(\pi)}_{ab}(\theta,\varphi_{p};\beta_I,\beta_F)
    &\equiv&\frac{\langle \beta_F| c_a^{(\pi)\dagger}c_b^{(\pi)\dagger}|
    \tilde{\beta_I}\rangle^{(\pi)}}{ \langle \beta_F
    |\tilde{\beta_I}\rangle^{(\pi)}},\\
    \kappa^{10(\nu)}_{cd}(\theta,\varphi_{n};\beta_I,\beta_F)&\equiv&
    \frac{\langle \beta_F| c_d^{(\nu)}c_c^{(\nu)}|
    \tilde{\beta_I}\rangle^{(\nu)}}{ \langle \beta_F
    |\tilde{\beta_I}\rangle^{(\nu)}}.~~~~~~~~~
\end{eqnarray}
\end{subequations}
Details about the evaluation of the two-body matrix element (TBME) $\langle ab|\hat{O}|cd\rangle$ in Eq.~(\ref{me}) is given in the next section and in the Appendix.

\section{Numerical Details}\label{numerical}

In the present work we restrict ourselves to axial symmetry. In this case the complicated GCM+PNP +3DAMP model is reduced to a relatively simple GCM+PNP+1DAMP calculation.

\begin{figure}[!htbp]
    \centering
    \includegraphics[width=7cm]{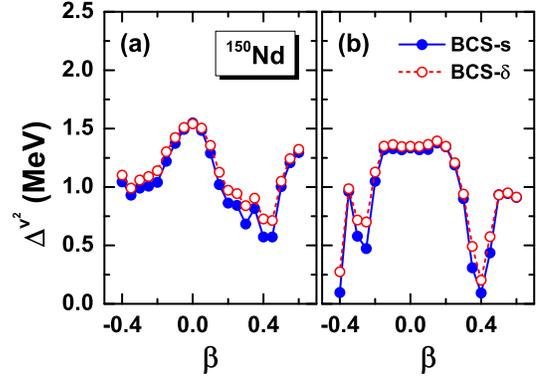}\\
    \caption{(Color online) Average pairing gap $\Delta^{v^2}$ for (a) neutrons and (b) protons in ${}^{150}$Nd as a function of deformation $\beta$ obtained by the RMF+BCS method, using the separable finite-range pairing force (BCS-s) and the $\delta$ pairing force with adjusted strength constants $V_\tau$ (BCS-$\delta$), respectively.}\label{fig01}
\end{figure}

On the mean-field level, to obtain the set of intrinsic states $|\beta\rangle$ with different deformations $\beta$, constrained RMF calculations are performed with the pair correlations treated by the BCS method. To solve the Dirac equation the single-particle states are expanded in the three-dimensional harmonic oscillator basis~\cite{Gambhir1990} with $N_\mathrm{sh}=12$ major shells.
We use the nonlinear point-coupling functional PC-PK1~\cite{Zhao2010} in the particle-hole channel, and the density-independent $\delta$ force in the particle-particle channel.
In particular, the pairing strength constants $V_\tau$ for neutrons and protons are adjusted by reproducing the average pairing gap,
\begin{eqnarray}
  \Delta^{v^2}\equiv \frac{\sum_k f_kv_k^2\Delta_k}{\sum_k f_kv_k^2},
\end{eqnarray}
provided by the separable finite-range pairing force~\cite{Tian2009_PLB, Tian2009_PhysRevC.80}, where $f_k=f(\varepsilon_k)$ is an energy-dependent cutoff function given in Ref.~\cite{Bender2000}. With the adopted values $V_n=-314.55~\mathrm{MeV~fm}^3$ and $V_p=-346.5~\mathrm{MeV~fm}^3$, the average pairing gaps are reproduced very well at different deformations, as shown in Fig.~\ref{fig01}.

In the PNP+1DAMP (PNAMP from now on) procedure, a Gaussian-Legendre quadrature is used for the integrals over the gauge angle $\varphi$ and the Euler angle $\theta$. Convergence of the potential energy curves (PECs) can be reached when the numbers of mesh points for $\varphi$ and $\theta$ in the interval $[0,\pi]$ are chosen to be $n_\varphi=7$ and $n_\theta=14$.

In the GCM calculation, the generator coordinates are chosen in the interval $\beta\in[-0.4,0.6]$ with a step size $\Delta\beta=0.1$. In the Hill-Wheeler-Griffin equation, eigenvectors of the norm overlap kernel with very small eigenvalues $n_k^J/n_\mathrm{max}^J<\chi$ are removed from the GCM basis~\cite{Yao2010}. For the chosen generator coordinates and the cutoff parameter $\chi=1\times 10^{-3}$, fully converged results can be achieved for the low-lying states with $J\leq6$ in ${}^{150}$Nd and ${}^{150}$Sm. Finally ten natural states are included for the $J=0$ states.

\begin{figure}[!htbp]
\centering
    \includegraphics[width=7cm]{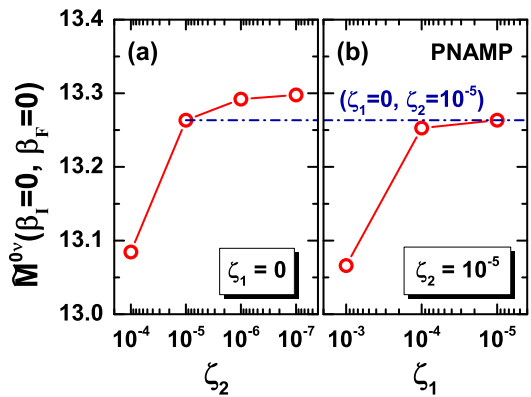}\\
\caption{(Color online) Single-configuration matrix element $\tilde{M}^{0\nu}(\beta_I, \beta_F)$ defined in Eq.~(\ref{MIFnorm}) between the spherical states of ${}^{150}$Nd and ${}^{150}$Sm, as a function of the cutoff parameters $\zeta_1$ and $\zeta_2$, respectively. The horizontal dash-dotted line denotes the value corresponding to $\zeta_1=0$ and $\zeta_2=10^{-5}$.}
\label{fig02}
\end{figure}

From the last section we see that we obtain the transition matrix element $M^{0\nu}(\beta_I,\beta_F)$ by evaluating expression (\ref{me}). As a basis we use for the large and small components of the single-particle spinors $|a\rangle$, $|b\rangle$, $|c\rangle$, $|d\rangle$ the spherical harmonic oscillator (SHO) states [for details, see Eq. (\ref{basis})]. In this case the following expression has to be calculated at every mesh point of the Euler angle $\theta$, the gauge angles $(\varphi_{n}, \varphi_{p})$, and the generator coordinates $(\beta_I, \beta_F)$:
\begin{eqnarray}
\label{sum}
    &&\sum_{1234}( 12|\hat{O}_{2\times 2}|34)
    ~n(\theta,\varphi_{n},\varphi_{p}; \beta_I, \beta_F)\\
    &&~~~~~~~\times \kappa_{12}^{01*(\pi)}(\theta,\varphi_{p}; \beta_I, \beta_F)
    \kappa_{43}^{10(\nu)}(\theta,\varphi_{n}; \beta_I, \beta_F).\notag
\end{eqnarray}
The notation $|1)$ refers to the SHO wave function $|1)\equiv|n_1l_1j_1m_1p_1\rangle$ with the radial quantum number $n$, the angular momentum quantum numbers $j,m$, and the quantum number $p=f,g$ characterizing large and small components of the relativistic spinor.  Because we express here the scalar product of the initial and final spinors explicitly, the operator $\hat{O}_{2\times 2}$, depending on $p_1$, $p_2$, $p_3$, and $p_4$, is part of the full $4\times4$ matrix $\hat O$ in Eq.~(\ref{me}). The summation $\sum_{1234}$ in Eq.~(\ref{sum}) includes a fourfold loop of the complete SHO basis. To reduce the computational effort we introduce additional cutoff parameters $\zeta_1$ and $\zeta_2$ to avoid in this loop the calculation of terms with small contributions:
\begin{eqnarray}
  \kappa_{12}^{01*(\pi)}< \zeta_1\quad \text{or} \quad \kappa_{12}^{01*(\pi)}\kappa_{43}^{10(\nu)} < \zeta_2.
\end{eqnarray}
In the case of spherical symmetry corresponding numerical checks have been carried out. In Fig.~\ref{fig02} we study the influence of the cutoff parameters on the single-configuration matrix elements $\tilde{M}^{0\nu}(\beta_I, \beta_F)$ defined in Eq.~(\ref{MIFnorm}). In the following applications we used the values of $\zeta_1=10^{-4}$ and $\zeta_2=10^{-5}$ with resulting errors less than $1\%$ for $\tilde{M}^{0\nu}(\beta_I, \beta_F)$, and with a considerable reduction of computer time.

\begin{figure}
  \centering
  \includegraphics[width=7cm]{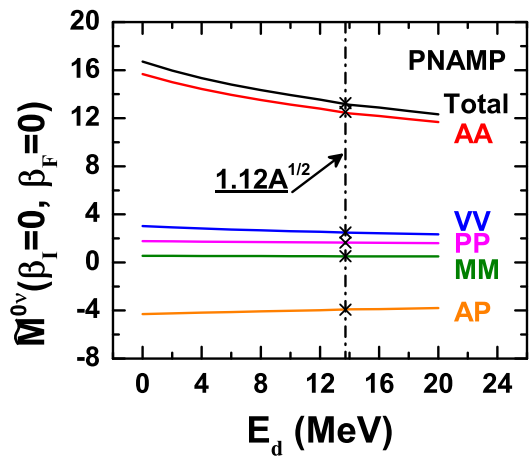}\\
  \caption{(Color online) Single-configuration matrix element $\tilde{M}^{0\nu}(\beta_I, \beta_F)$ between the spherical states of ${}^{150}$Nd and ${}^{150}$Sm, as a function of
  the energy denominator $E_d$ in Eq.~(\ref{operator1}). The empirical value of $E_d=1.12A^{1/2}~\mathrm{MeV}$ is marked by a vertical dash-dotted line.}\label{fig03}
\end{figure}

At last, the reliability of the closure approximation has to be tested in the relativistic scenario. To that end, we change the values of $E_d$ in Eq.~(\ref{operator1}) from $0$ to $20~\mathrm{MeV}$ and compare the corresponding single-configuration matrix element $\tilde{M}^{0\nu}(\beta_I=0, \beta_F=0)$. In Fig.~\ref{fig03} it is shown that the matrix element and the contributions from different channels are insensitive to the change of $E_d$. In particular, the calculations with $8~\mathrm{MeV}\leq E_d\leq 20~\mathrm{MeV}$ lead to similar values for the matrix element with derivations less than $10\%$ from its central value. The empirical value $E_d=1.12A^{1/2}~\mathrm{MeV}\simeq 13.72~\mathrm{MeV}$ proposed by Haxton \emph{et al.}~\cite{Haxton1984} is used in the present calculations. This is very close to the central value we just mentioned.

\section{Results and Discussion}\label{result}

\subsection{Nuclear structure properties}

\begin{figure}[!htbp]
    \centering
    \includegraphics[width=8.5cm]{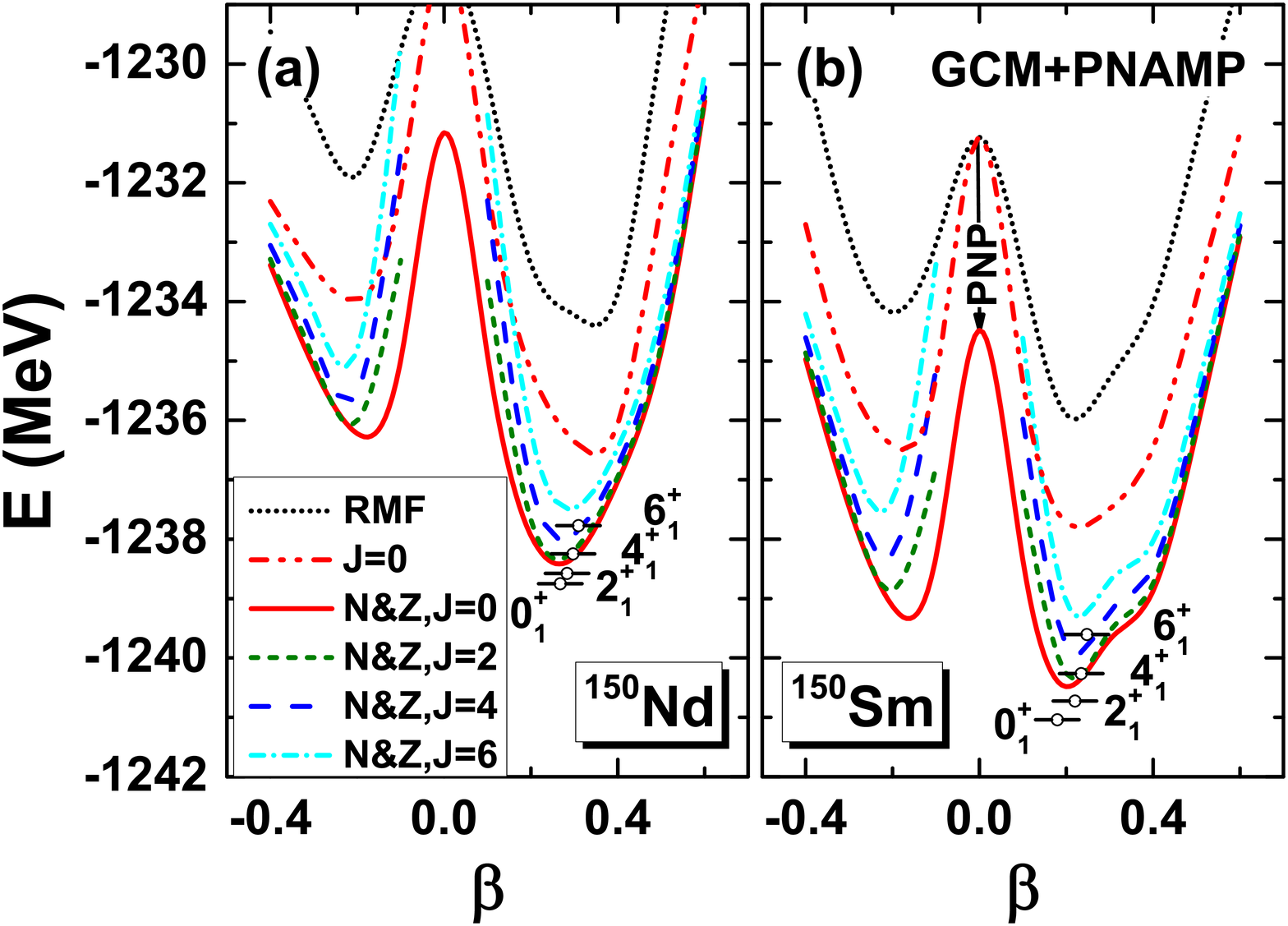}\\
    \caption{(Color online) The intrinsic (RMF) and the PNAMP ($N\&Z, J=0, 2,4,6$) PECs, together with the energy and the average axial deformation of the lowest GCM state for each angular momentum in ${}^{150}$Nd and ${}^{150}$Sm. The AMP PECs, which are provided by calculations without exact number projection, are also presented for $J=0$.}\label{fig04}
\end{figure}

The GCM+PNAMP calculations have been carried out to obtain the wave functions for the initial and final states used in the evaluation of the NMEs for the $0\nu\beta\beta$ decay. In Fig.~\ref{fig04} the intrinsic PECs are shown derived from constrained RMF+BCS calculations for the nuclei ${}^{150}$Nd and ${}^{150}$Sm, as well as the corresponding angular momentum and particle number projected PECs with $J=0,2,4,6$. For $\beta=0$ the AMP has no influence. The lowering in energy at this point is therefore caused only by number projection. For both nuclei we observe energy gains of $2\sim 5~\mathrm{MeV}$ by the number projection. A prolate deformed minimum and an oblate deformed local minimum are observed for each of the PECs. For ${}^{150}$Nd the unprojected prolate minimum is rather flat. In fact, as observed in experiment~\cite{Kruecken2002} and also found in GCM calculations~\cite{Niksic2007} based on the PC-F1, this nucleus is very close to a quantum phase transition from spherical to prolate with a spectrum of X(5) character~\cite{Iachello2001}. Therefore, it is essential to take into account for this nucleus quantum fluctuations in deformation space. For both nuclei rotational yrast bands are constructed by AMP after the variation based on the wave functions around the prolate minimum, with average axial deformations $\beta\simeq0.3$ for ${}^{150}$Nd and $\beta\simeq0.2$ for ${}^{150}$Sm.

In Fig.~\ref{fig04}, the angular momentum projected energy curves (without PNP) of $J=0$ with the average particle numbers constrained~\cite{Yao2010, Yao2011} are also included (dash-double-dotted line). By comparison one can see that the exact PNP shifts the position of the energy minimum for ${}^{150}$Nd to smaller deformation. This could be possibly understood by the fact that PNP increases slightly the pairing correlations driving to smaller deformations.

\begin{figure}[!htbp]
    \centering
    \includegraphics[width=7cm]{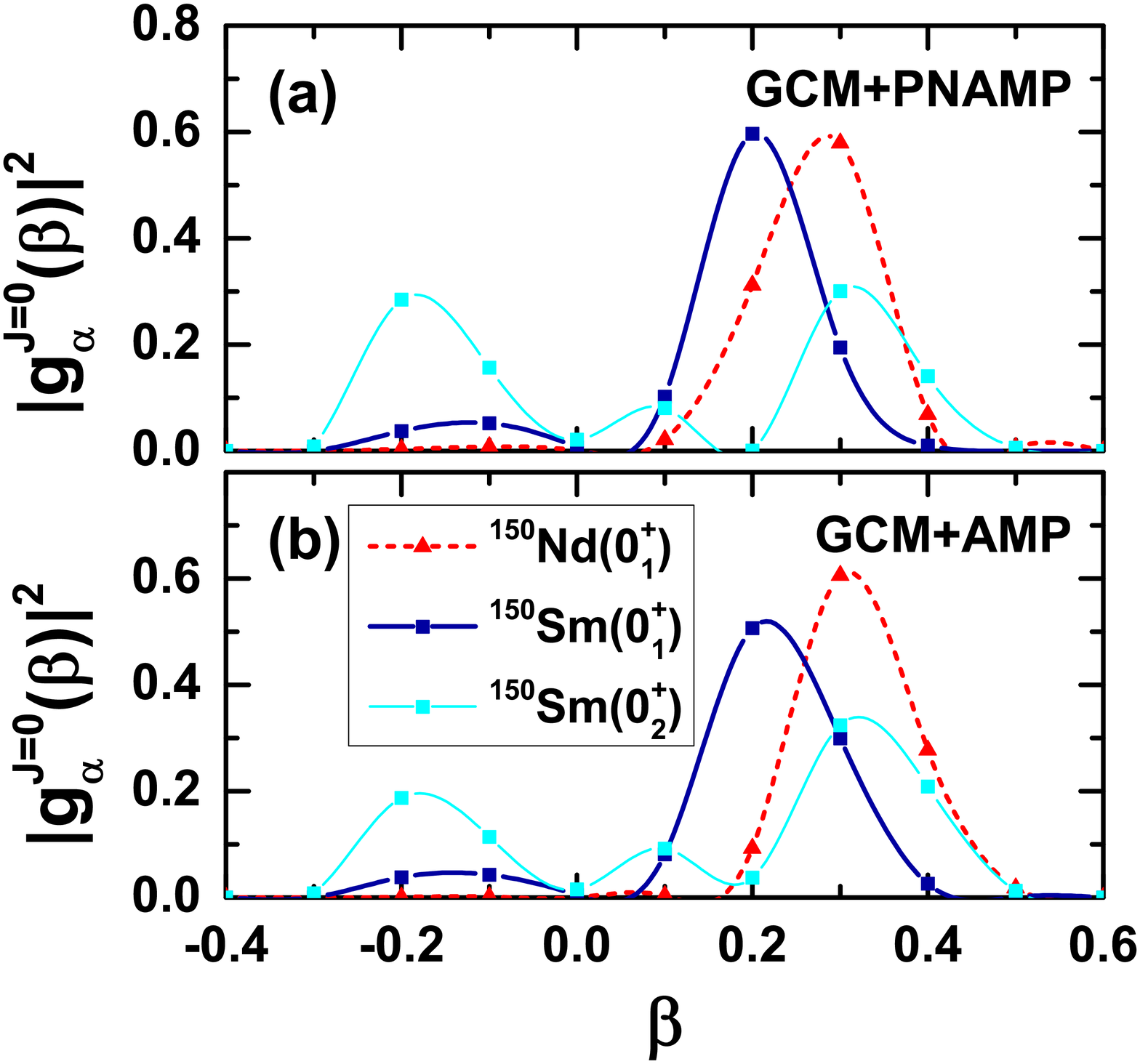}\\
    \caption{(Color online) Squares of collective wave functions $|g_\alpha^{J=0}(\beta)|^2$ obtained by the GCM+PNAMP and GCM+AMP methods for the ground states of ${}^{150}$Nd and ${}^{150}$Sm, as well as for the first excited $0^+$ state of ${}^{150}$Sm.}\label{fig05}
\end{figure}

In Fig.~\ref{fig05} we show the squares of collective wave functions defined in Eq.~(\ref{gfunction}) for the $0^+$ states, which denote the probability distributions of the corresponding states in deformation space. For the ground state of ${}^{150}$Nd, wave functions calculated by both the GCM+PNAMP and the GCM+AMP methods are peaked at $\beta=0.3$, but the probability distribution shifts from the right side of the peak with larger deformation to the left side with weaker deformation after considering the PNP. The change in collective wave functions of this nucleus is consistent with the change of shapes of the $J=0$ energy curve observed in Fig.~\ref{fig04}(a) with and without PNP. Meanwhile, the wave functions of the $0_1^+$ and $0_2^+$ states of ${}^{150}$Sm obtained by the two methods are very similar. Consequently, the overlap between ${}^{150}\mathrm{Nd}(0_1^+)$ and ${}^{150}\mathrm{Sm}(0_1^+)$ increases by PNP, while the overlap between ${}^{150}\mathrm{Nd}(0_1^+)$ and ${}^{150}\mathrm{Sm}(0_2^+)$ decreases.

\begin{figure}[!htbp]
    \centering
    \includegraphics[width=7.5cm]{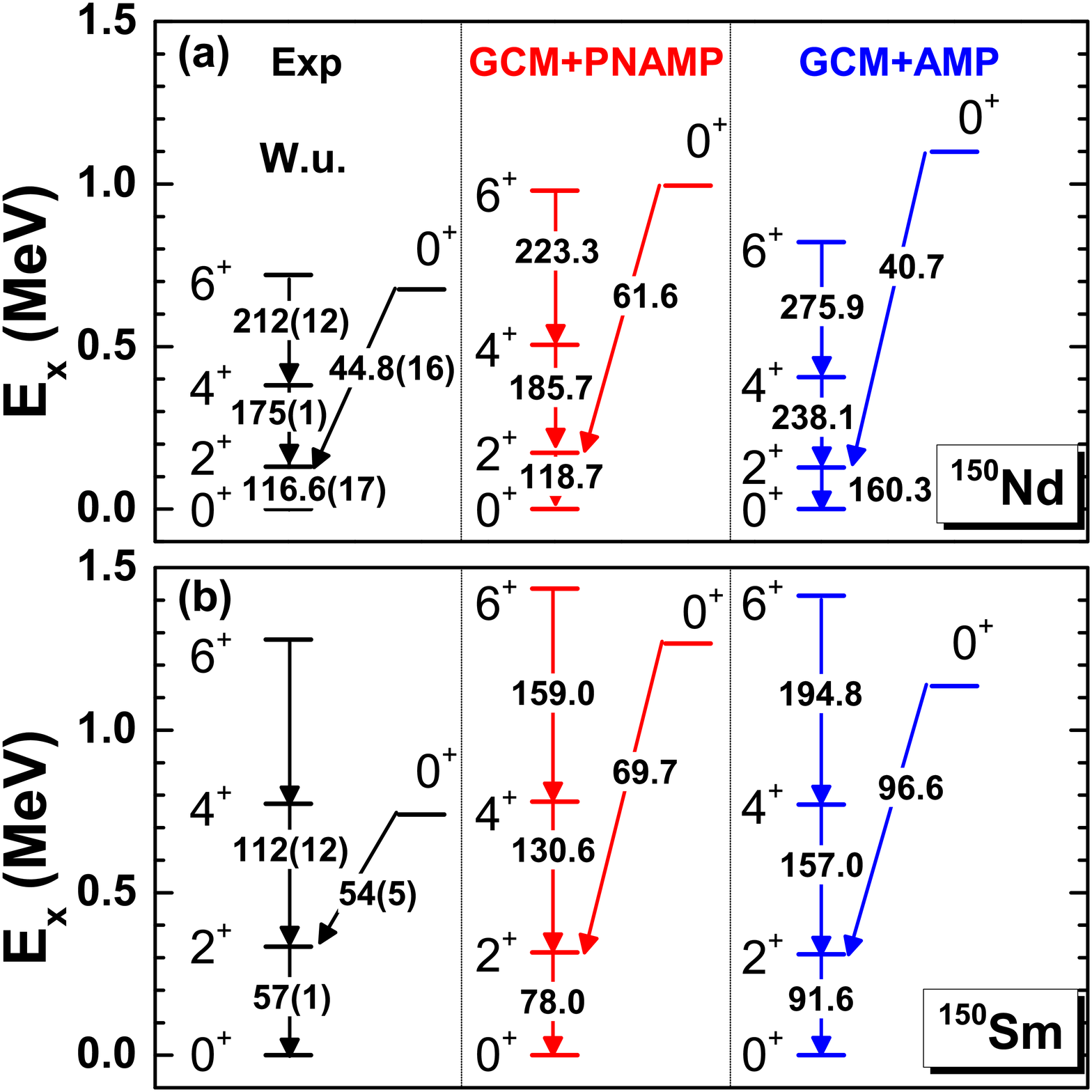}\\
    \caption{(Color online) Low-lying energy levels and $E2$ transition probabilities for the nuclei ${}^{150}$Nd and ${}^{150}$Sm obtained by the GCM+PNAMP and GCM+AMP methods in comparison with experimental data.}\label{fig06}
\end{figure}

To prove the validity of our model for the description of ${}^{150}$Nd and ${}^{150}$Sm, we show in Fig.~\ref{fig06} their low-lying excitation properties obtained by the GCM+PNAMP and GCM+AMP methods and compare them with available experimental data. It turns out that the GCM+AMP calculation reveals similar characteristics as the GCM+PNAMP method. The level schemes are in rather good agreement with the data, but in both cases the calculated spectra are systematically stretched as compared to the experimental bands. This is a well-known fact observed also in other calculations of this type~\cite{Yao2014}: Because AMP is performed only after variation, time-odd components and alignment effects are neglected, leading to an underestimated momentum of inertia. The agreement of the calculated $E2$ transition probabilities with data is remarkable, especially in the case of GCM+PNAMP. This indicates that our GCM+PNAMP-wave functions have very good deformation properties as compared to experiment.

\subsection{Nuclear matrix elements}
\subsubsection{Effects of number projection}

To check the numerical accuracy of our projection techniques, we investigate the relation (\ref{singlePNP}) numerically; i.e., we show that single PNP is equivalent to double PNP in the calculation of the matrix element for the $0\nu\beta\beta$ decay operator.

In Table~\ref{tab01}, $n_{\varphi_I}$ ($n_{\varphi_F}$) denotes the number of mesh points used in the integrals (\ref{me}) over the gauge angle in the neutron or proton number projection for the initial (final) state. The calculation reduces to the pure AMP case when the number of mesh points is set to $1$. As shown in the table, for the matrix elements of $\hat{\mathcal O}^{0\nu}$, calculations with single PNP for the initial state, with single PNP for the final state, and with double PNP for both of the states lead, as expected, to identical results. This shows clearly that number projection is carried out with sufficient accuracy in our calculations. Therefore, in practice, we only keep the projection operators on the side of the mother nucleus.

\begin{table}
  \centering
  \caption{Matrix elements of the $0\nu\beta\beta$ decay operator $\langle\beta_F|\hat P^{N_F}\hat P^{Z_F}\hat{\mathcal O}^{0\nu}\hat P^{J=0}\hat P^{N_I}\hat P^{Z_I}|\beta_I\rangle$ and contributions from the various coupling channels. The results without PNP ($n_{\varphi_I}=1$, $n_{\varphi_F}=1$), with single PNP for the initial state ($n_{\varphi_I}=7$, $n_{\varphi_F}=1$), with single PNP for the final state ($n_{\varphi_I}=1$, $n_{\varphi_F}=7$), and the results with double PNP ($n_{\varphi_I}=7$, $n_{\varphi_F}=7$) are compared.}\label{tab01}
  \begin{tabular}{cc|cccccc}
    \hline\hline
     ~$n_{\varphi_I}$~&~$n_{\varphi_F}$~& ~~VV~~ & ~~AA~~ & ~~AP~~ & ~~PP~~ & ~~MM~~ & ~~Total~~ \\\hline
     $1$ & $1$ & $2.552$ & $12.588$ & $-4.025$ & $1.698$ & $0.519$ & $13.332$ \\
     $7$ & $1$ & $0.196$ & $0.982$  & $-0.309$ & $0.130$ & $0.040$ & $1.039$  \\
     $1$ & $7$ & $0.196$ & $0.982$  & $-0.309$ & $0.130$ & $0.040$ & $1.039$  \\
     $7$ & $7$ & $0.196$ & $0.982$  & $-0.309$ & $0.130$ & $0.040$ & $1.039$  \\
    \hline\hline
  \end{tabular}
\end{table}

\begin{figure}[!htbp]
    \centering
    \includegraphics[width=7cm]{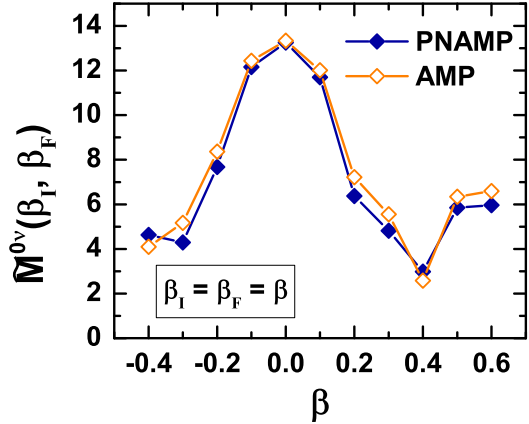}\\
    \caption{(Color online) Single-configuration matrix elements $\tilde{M}^{0\nu}(\beta_I, \beta_F)$ defined in Eq.~(\ref{MIFnorm}) with $\beta_I=\beta_F$ for transitions from ${}^{150}$Nd to ${}^{150}$Sm, obtained by calculations with PNP (PNAMP) and without (AMP).}\label{fig07}
\end{figure}

To investigate the effect of number projection on the $0\nu\beta\beta$ decay matrix elements, we display in Fig.~\ref{fig07} the values of single-configuration matrix elements $\tilde{M}^{0\nu}(\beta_I, \beta_F)$ in Eq.~(\ref{MIFnorm}) obtained with and without PNP in the case of $\beta_I=\beta_F$. As we can see, for both the spherical and the deformed cases, the values of the single-configuration matrix elements are not significantly affected by PNP. Of course, this applies only for the matrix elements with fixed deformation. However, as we see in Fig.~\ref{fig05}, the weights of the different deformations in the GCM wave functions depend on PNP and therefore, when using the full GCM matrix elements, one should include PNP.

\subsubsection{Effects of deformation}

\begin{figure}[!htbp]
    \centering
    \includegraphics[width=8.5cm]{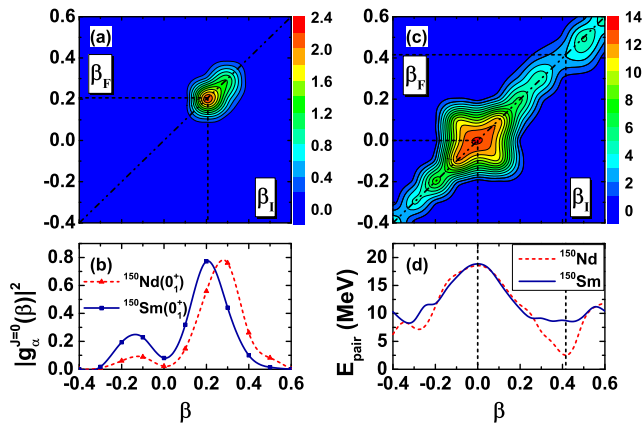}\\
    \caption{(Color online) (a) Distributions of the total transition matrix element $M^{0\nu}$ of Eq.~(\ref{M}) between the ground states of ${}^{150}$Nd and ${}^{150}$Sm in the various regions of the  $\beta_I$-$\beta_F$ plane, calculated with the GCM+PNAMP method and (c) normalized matrix element $\tilde{M}^{0\nu}(\beta_I, \beta_F)$ of Eq.~(\ref{MIFnorm}) obtained by the single-configuration calculation with PNAMP. (b) Squares of ground-state wave functions obtained with the GCM+PNAMP method and (d) pairing energies (\ref{Epair}) from the RMF+BCS calculation for initial and final nuclei are shown for comparison.}\label{fig08}
\end{figure}

The NME $M^{0\nu}$ in Eq.~(\ref{M}) is a superposition of un-normalized matrix elements $M^{0\nu}(\beta_I,\beta_F)$ with various deformations $(\beta_I, \beta_F)$ multiplied with specific weights. From Eq. (\ref{M}) it is evident that configuration mixing occurs and that the regions of maximal overlap between the three quantities $f^\ast_{0_F^+}(\beta_F)$, $f_{0_I^+}(\beta_I)$, and $M^{0\nu}(\beta_I,\beta_F)$ contribute mostly to the total matrix element $M^{0\nu}$. In Fig.~\ref{fig08}, the distribution of $f^\ast_{0_F^+}(\beta_F)f_{0_I^+}(\beta_I)M^{0\nu}(\beta_I,\beta_F)$ is displayed for the transition between ${}^{150}\mathrm{Nd}(0_1^+)$ and ${}^{150}\mathrm{Sm}(0_1^+)$ in panel (a). Therefore, this figure shows which configurations contribute dominantly in the $\beta_I$-$\beta_F$ plane. As we can see in Fig.~\ref{fig08}(a) the largest contributions come from the region $\beta_I\simeq\beta_F\simeq0.2$. Similar deformation of the initial and final states is favored by the decay process. Therefore, a large overlap between the initial and the final collective wave functions is important. In Fig.~\ref{fig08}(b) we show the collective wave functions for the ground states of the two nuclei as a function of the deformation. It is clearly seen, that these distributions are peaked at $\beta\simeq0.3$ for the nucleus ${}^{150}$Nd and at $\beta\simeq0.2$ for the nucleus ${}^{150}$Sm. However, the distributions show a relatively large width and therefore there is an overlapping region of considerable size in between. It is evident that deformation fluctuation plays an essential role in the description of the transition matrix element.

The situation is rather different when we consider the normalized single-configuration matrix element $\tilde{M}^{0\nu}(\beta_I, \beta_F)$ defined in Eq. (\ref{MIFnorm}). This matrix element is shown in Fig.~\ref{fig08}(c) as a function of the initial and final deformations $\beta_I$ and $\beta_F$. It is no longer related to collective wave functions; rather it is assumed that the initial nucleus has a fixed intrinsic deformation $\beta_I$ and the final nucleus has another intrinsic deformation $\beta_F$. The value of the matrix element is then taken from the corresponding point in Fig.~\ref{fig08}(c). Obviously, this method provides a reasonable approximation only for transitions between nuclei with well-defined intrinsic deformations, i.e., sharp minima in the PECs and narrow collective wave functions.

Figure~\ref{fig08}(c) shows that the single-configuration matrix element is peaked at zero deformation. This fact is consistent with the previous nonrelativistic GCM+PNAMP calculations of Ref.~\cite{Rodriguez2010}.  It can be understood by the fact that the expression given in Eq. (\ref{sum}) has in the diagonal case a structure similar to that of the pairing energy
\begin{equation}
\label{Epair}
E_{pair}(\beta)=\frac12\sum_{1234}\langle 12|V^{pp}|34\rangle\kappa_{12}(\beta)\kappa_{43}(\beta),
\end{equation}
where $V^{pp}$ is the effective pairing interaction in the $pp$ channel. Therefore, a strong correlation can be found between $\tilde{M}^{0\nu}(\beta_I, \beta_F)$ and the pairing correlations. It is well known that minima in the PEC are strongly connected with low level densities and weak pairing, whereas maxima in the PEC are connected with high level densities and strong pairing correlations. Therefore, we have at zero deformation enhanced pairing energies and enhanced transition matrix elements $\tilde{M}^{0\nu}(\beta_I, \beta_F)$. Similar effects have been observed in double humped fission barriers~\cite{Karatzikos2010}. Figure~\ref{fig08}(d) shows the pairing energy as a function of the deformation. We have to keep in mind, however, that the strongly enhanced transition matrix elements at small deformation have little to do with the $0\nu\beta\beta$ decay matrix element between the ground states of the nuclei ${}^{150}$Nd and ${}^{150}$Sm with strong intrinsic deformations.

In Table~\ref{tab02} we show the influence of correlations owing to projections and of fluctuations treated in GCM on the $0\nu\beta\beta$ matrix elements. In the second column we show single-configuration matrix elements with and without change of the intrinsic deformation. These NMEs $\tilde{M}^{0\nu}(\beta_I,\beta_F)$ with $\beta_F\neq\beta_I$ are given at the deformations corresponding to the minima on the $J^\pi=0^+$ energy surfaces of ${}^{150}$Nd and ${}^{150}$Sm. We observe that AMP enhances the NMEs and additional number projection reduces them. Also listed are NMEs neglecting the change of deformation ($\beta_F=\beta_I$). They are considerably larger, because it is well known that the many-body overlap functions $\langle \beta|\hat{O}|\beta'\rangle$ are sharply peaked at $\beta=\beta'$. In the third column fluctuations are taken into account in the framework of the GCM method. As discussed in the last paragraph this enhances the transition matrix elements, compared to the matrix element between energy minima (the $\beta_F\neq \beta_I$ case), because of the enhanced overlap owing to the width in the collective wave functions [see Fig.~\ref{fig08}(b)]. In this case PNP leads to an additional increase of the transition matrix element $M^{0\nu}$, because, as shown in Fig.~\ref{fig05}, the changes in the collective wave functions induced by PNP lead to an enhanced larger overlap.
\begin{table}[!htbp]
	\centering
\caption{NMEs for the $0\nu\beta\beta$ decay between ${}^{150}$Nd and ${}^{150}$Sm, with different correlations considered in the nuclear ground-state wave functions. Single-configuration matrix elements in the second column are compared with GCM results in the third column.}
	\begin{tabular}{l|cc|c}
	      \hline\hline
	      &\multicolumn{2}{c|}{~~~$\tilde M^{0\nu}(\beta_I,\beta_F)$~~~} &  ~$M^{0\nu}(0^+_1\rightarrow 0_1^+)$~ \\
	      \cmidrule(r){2-3}
	      & ~$\beta_F\neq\beta_I$~  & ~$\beta_F=\beta_I$~ &  \\\hline
	      ~BCS                 & $3.56$    & $6.38$ &            \\
	      ~AMP                 & $3.88$    & $6.79$ &            \\
	      ~PNAMP            & $3.27$    & $6.02$ &            \\\hline
	      ~GCM+AMP      &                &             & $4.68$ \\
	      ~GCM+PNAMP~ &                &             & $5.60$ \\
	      \hline\hline
	\end{tabular}
\label{tab02}
\end{table}

Summarizing this section, we see that in transitional nuclei the $0\nu\beta\beta$ decay matrix elements depend in a sensitive way on the deformation and on the pairing properties of these nuclei, which are taken into account with different accuracy in the various methods. The details depend much on the nucleus under consideration. GCM+PNAMP is, of course, the most appropriate method. It could be possibly further improved in specific nuclei with triaxial deformations by 3D AMP and 2D GCM in the ($\beta,\gamma$) plane. This, however, leads in medium-heavy and heavy nuclei to considerable numerical efforts at the limit of the present days' computer facilities~\cite{Yao2014}. As shown in Ref.~\cite{Yao2014}, investigations of nuclear spectra calculations within microscopic versions of the 5D-collective Bohr Hamiltonian provide a very successful alternative which can be applied even in heavy nuclei~\cite{Li2010_PhysRevC.81}. It remains to be investigated in the future whether these methods can be used also successfully for studies of the $0\nu\beta\beta$ decay matrix elements.

\subsubsection{Validity of nonrelativistic reduced calculations and contribution of the tensor term}

One advantage of our method is that it is fully relativistic and therefore it allows us to investigate the nonrelativistic approximation applied in most calculations. In this case, the hadronic current $\mathcal J_\mu^\dagger(\bm x)$ in Eq. (\ref{operator1}) is expanded in terms of $|\bm q|/m_p$. If terms are kept up to the first order, the fully relativistic operator of Eq. (\ref{operator1}) is reduced to the nonrelativistic operator used in previous studies~\cite{Ericson1988, Simkovic2008}. The resulting nonrelativistic ``two-current'' operator $\left[\mathcal J_\mu^\dagger\mathcal J^{\mu\dagger}\right]_\mathrm{NR}$ can be decomposed, as in other nonrelativistic calculations, into the Fermi, the Gamow-Teller, and the tensor parts,
\begin{eqnarray}\label{hq}
\left[-h_{F}(\bm q^2)+ h_{GT}(\bm q^2)\sigma_{12}+h_{T}(\bm q^2)S^q_{12}\right]\tau_-^{(1)}\tau_-^{(2)},
\end{eqnarray}
with the tensor operator $S^q_{12}=3(\bm \sigma^{(1)}\cdot\hat{\bm q})(\bm \sigma^{(2)}\cdot\hat{\bm q})-\sigma_{12}$ and $\sigma_{12}=\bm\sigma^{(1)}\cdot\bm\sigma^{(2)}$. Each channel ($K: F, GT, T$) of Eq.~(\ref{hq}) can be labeled by the terms of the hadronic current from which it originates, as
\begin{equation*}
    h_K(\bm q^2)=\sum_i h_{K-i}(\bm q^2),\quad(i=VV,AA,AP,PP,MM)
\end{equation*}
with
\begin{subequations}
	\begin{eqnarray}\label{twocurrentNR}
		h_{{F}-VV}(\bm q^2) &=&-g_V^2(\bm q^2),\\\label{FVV}
	    h_{{GT}-AA}(\bm q^2)&=&-g_A^2(\bm q^2),\\\label{GTAA}
		h_{{GT}-AP}(\bm q^2)&=&~~\frac23g_A(\bm q^2)g_P(\bm q^2)\frac{\bm q^2}{2m_p},\\\label{GTAP}
		h_{{GT}-PP}(\bm q^2)&=&-\frac13g_P^2(\bm q^2)\frac{\bm q^4}{4m_p^2},\\\label{GTPP}
		h_{{GT}-MM}(\bm q^2)&=&-\frac23g_M^2(\bm q^2)\frac{\bm q^2}{4m_p^2},\\\label{GTMM}
        h_{{T}-AP}(\bm q^2)&=&~~h_{GT-AP}(\bm q^2),\\\label{TAP}
		h_{{T}-PP}(\bm q^2)&=&~~h_{GT-PP}(\bm q^2),\\\label{PP}
		h_{{T}-MM}(\bm q^2)&=&-\frac12h_{GT-MM}(\bm q^2).\label{TMM}
	\end{eqnarray}
\end{subequations}

\begin{figure}[!htbp]
    \centering
    \includegraphics[width=8cm]{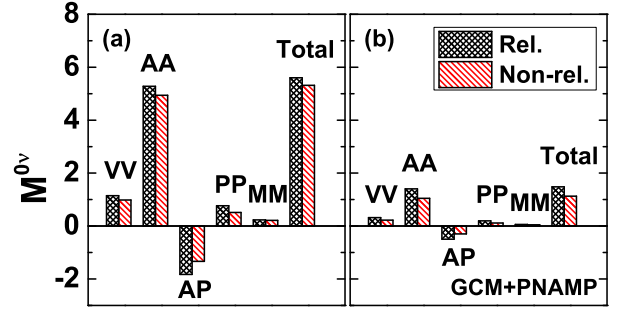}\\
    \caption{(Color online) Contribution from each coupling channel to the total NMEs of the $0\nu\beta\beta$ decay from ${}^{150}$Nd to ${}^{150}$Sm for both the (a) $0_1^+\rightarrow 0_1^+$ and the (b) $0_1^+\rightarrow 0_2^+$ transitions. Values of the matrix elements evaluated using the full relativistic operator $M^{0\nu}$ (Rel.) are compared with those obtained with the nonrelativistic reduced operator $M_\text{NR}^{0\nu}$ (Non-rel.). The results are calculated with the GCM+PNAMP method.}\label{fig09}
\end{figure}

In Fig.~\ref{fig09} we compare the results calculated with the nonrelativistic reduced operator with those of the full operator, for the NME in each coupling channel, and for both the $0_1^+\rightarrow 0_1^+$ and $0_1^+\rightarrow 0_2^+$ transitions. 
In all circumstances the dominant contributions come from the AA coupling channel. In the nonrelativistic approximation it represents the Gamow-Teller channel if neglecting the high-order currents. In this comparison, the relativistic effect $\Delta_\text{Rel.}\equiv(M^{0\nu}-M^{0\nu}_\text{NR})/M^{0\nu}$ is roughly $5\%$ in the $0_1^+\rightarrow 0_1^+$ transition and $24\%$ in the $0_1^+\rightarrow 0_2^+$ transition.

We divide our GCM+PNAMP NME obtained with the nonrelativistic operator into the Gamow-Teller, the Fermi, and the tensor matrix elements, as $M_\mathrm{NR}^{0\nu} =M_{GT} -M_{F} +M_{T}$, and show for the $0_1^+\rightarrow 0_1^+$ transition in Table~\ref{tab03}. They are compared with the NREDF results~\cite{Rodriguez2010} and the IBM-2 calculations~\cite{Barea2013}. Note that the definition of the Fermi matrix element $M_{F}$ is different from Eq.~(19) in Ref.~\cite{Barea2013} by a factor of $\left(g_V(0)/g_A(0)\right)^2$. Considering $\chi_{F}=-{M_{F}}/{M_{GT}}$ and $\chi_{T}={M_{T}}/{M_{GT}}$, the ratios of the Fermi and tensor parts to the dominant Gamow-Teller part, one clearly recognizes the contributions of the these terms.

It is shown that the Fermi contribution ($33.6\%$) in the NREDF calculation is relatively large compared to our results, while the IBM-2 model gives a much smaller value ($8.9\%$). As a matter of fact, the IBM-2 calculations provide very small Fermi matrix elements for the nuclei in which protons and neutrons occupy different major shells (for example, ${}^{150}$Nd-Sm), and very large values for those in which protons and neutrons occupy the same major shell (for example, ${}^{76}$Ge-Se)~\cite{Barea2013}. A benchmark study is definitely required to understand the discrepancy among different models in the future.
However, it has been pointed out in Ref.~\cite{Simkovic2013} that, with partial isospin symmetry restoration by requiring $M_{F}^{2\nu}=0$, the value of $\chi_F$ for the matrix elements of neutrinoless double-$\beta$ decay should be close to $1/(3g^2_A(0))$. We find that our result ($23.5\%$) is in good agreement with the value of $1/(3g^2_A(0))=21\%$.

In the literature one finds rarely discussions about the tensor effect for the case of ${}^{150}$Nd. However, by analyzing the results for other isotopes, two different conclusions can be drawn. On the one hand, the tensor effect is considered as negligible according to the calculations in the ISM~\cite{Menendez2009} and PHFB~\cite{Rath2013}, and in the QRPA studies of the Jyv\"askyl\"a group~\cite{Kortelainen2007_PhysRevC.76}, and it is totally neglected in the NREDF calculations of Refs.~\cite{Rodriguez2010, Vaquero2013}. On the other hand, it is proven to be important with considerable contributions in the QRPA calculations of the T{\"u}bingen group~\cite{Simkovic1999} and in the IBM calculations~\cite{Barea2013}. Our result seems to agree with the later opinion. As we can see from the table, while the absolute value for the tensor term in our calculation is very close to that given by the IBM-2, $\chi_{T}$ is smaller owing to the larger Gamow-Teller contribution. This implies that we predict a relatively small tensor effect, but in the same order of magnitude as the IBM-2 calculations~\cite{Barea2013}.

\begin{table}[!htbp]
\centering
\caption{NMEs for the $0\nu\beta\beta$ decay between the ground states of ${}^{150}$Nd and ${}^{150}$Sm based on the nonrelativistic reduced operators, including the contributions of the Gamow-Teller, Fermi, and tensor terms. Our results with the GCM+PNAMP methods (REDF-I) are compared to the NMEs given by the NREDF calculation~\cite{Rodriguez2010} and the IBM-2 model~\cite{Barea2013}.}\label{tab03}
\begin{tabular}{l|ccccccr}
       \hline\hline
       &~~$M_\mathrm{NR}^{0\nu}$~~  & ~$M_{GT}$~ & ~$M_{F}$~ & ~$M_{T}$~~ & ~$\chi_{F}(\%)$~ & ~$\chi_{T}(\%)$~ \\
       \hline
       ~REDF-I~~   & $5.32$  & $4.22$  & $-0.99$  & $0.11$  & $23.5$ & $2.6$ \\
       ~NREDF~~ & $1.71$  & $1.28$  & $-0.43$  &    $-$          & $33.6$ & $-$ \\
       ~IBM-2~~            & $2.32$  & $2.03$  & $-0.18$  & $0.11$  & $8.9$ & $5.4$ \\
       \hline\hline
\end{tabular}
\end{table}

\subsubsection{Comparison and discussion}

\begin{table*}[!htbp]
    \centering
    \caption{NMEs for the $0\nu\beta\beta$ decay from ${}^{150}$Nd to ${}^{150}$Sm evaluated with different models. Results of this work are obtained with the GCM+PNAMP (REDF-I) and the GCM+AMP (REDF-II) methods. Also shown are the corresponding half-lives $T^{0\nu}_{1/2}$ for an assumed effective Majorana neutrino mass $\langle m_\nu\rangle=50~\mathrm{meV}$.}\label{tab04}
    \begin{tabular}{c|cccccc}
        \hline\hline
        & ~~~REDF-I~~~ & ~~~REDF-II~~~ & ~~NREDF\cite{Rodriguez2010, Vaquero2013}~~~ & ~~QRPA\cite{Fang2010,Mustonen2013}~~~ & ~~IBM-2\cite{Barea2013}~~~ & ~~PHFB\cite{Rath2013}~ \\
        \hline
        ~$M^{0\nu}(0_1^+\rightarrow 0_1^+)~~~~~~~~~~~~$                & $5.60$  & $4.68$   & $1.71,~~~2.19$ & $3.16,~~~2.71$ & $ 2.321$  & $2.83$ \\
        ~$T^{0\nu}_{1/2}(0_1^+\rightarrow 0_1^+)~[10^{25}~\text{y}]$  & $2.1~~$ & $3.1~~$  & $22.9,~~14.0~$ & $6.7~,~~~9.1~$ & $12.4~~$  & $8.4~$ \\
        ~$M^{0\nu}(0_1^+\rightarrow 0_2^+)~~~~~~~~~~~~$                & $1.48$  & $2.42$   & $2.81$\cite{Beller2013},~$-$     & $-$         & $0.395$   & $-$ \\
        ~$T^{0\nu}_{1/2}(0_1^+\rightarrow 0_2^+)~[10^{25}~\text{y}]$  &$70.7~~~$&$26.4~~$  & $19.6,~-$     & $-$        & $992.7~~$ & $-$ \\
        \hline\hline
    \end{tabular}
\end{table*}

In Table~\ref{tab04} we show the presently calculated $0\nu\beta\beta$ decay matrix elements $M^{0\nu}$ from ${}^{150}$Nd to ${}^{150}$Sm. The calculations are carried out in the MR-CDFT framework with the GCM+(PN)AMP method based on the REDF PC-PK1. These results are compared with existing results that take into account the nuclear deformations explicitly.

By taking into account nuclear deformations and configuration mixing simultaneously, we find in our calculation a suppression of approximately $60\%$ with respect to the spherical NME. The difference between the NMEs obtained with and without PNP (columns 2 and 3) can be traced back to differences in the distributions of the collective wave functions. As we have mentioned, the overlap between ${}^{150}\mathrm{Nd}(0_1^+)$ and ${}^{150}\mathrm{Sm}(0_1^+)$ is increased by PNP, resulting in a larger value of the matrix element $M^{0\nu}$ between them. The opposite holds for the matrix element $M^{0\nu}$ between ${}^{150}\mathrm{Nd}(0_1^+)$ and ${}^{150}\mathrm{Sm}(0_2^+)$.

NMEs obtained by the deformed QRPA calculations based on a Woods-Saxon field with a realistic residual interaction (the Brueckner $G$ matrix derived from the Bonn-CD potential)~\cite{Fang2010} can be found in column 5 of Table~\ref{tab04}. These matrix elements are suppressed by about $40\%$ by including the nuclear deformations as compared with the previous spherical QRPA results in Refs.~\cite{Rodin2006, Rodin2007}. More recently, a self-consistent Skyrme-HFB-QRPA calculation was carried out in Ref.~\cite{Mustonen2013}. It allows for an axially symmetric deformation and uses a modern Skyrme functional for both the HFB mean field and the QRPA. This investigation predicts a relatively small NME, which is also listed in column 5.

Calculations within the IBM model in Ref.~\cite{Barea2009, Barea2013} provide not only the NME for the transition to the ground state, but also for the transition to the first $0^+$ excited state. The IBM-2 interaction is used and the NME corresponding to the $0_1^+\rightarrow 0_1^+$ decay is $2.321$ (column 6). The inclusion of deformation causes only a reduction of about $20\%$~\cite{Barea2009}.

The recent result from the PHFB model~\cite{Rath2013} with a pairing plus quadrupole-quadrupole (PQQ) interaction is presented in column 7. Here the QQ term is responsible for the nuclear deformation.

A GCM calculation with projection has been recently carried out in the framework of the NREDF of Gogny D1S in Ref.~\cite{Rodriguez2010}. The concept is similar to ours. By choosing the deformation $\beta$ as the generator coordinate in the GCM method, the final NME includes the shape mixing effect and the resulting NME is $M^{0\nu}=1.71$ (column 4). Compared to the spherical case, this value is highly suppressed by more than $85\%$. NME for the transition to the $0_2^+$ state of ${}^{150}$Sm given by the same approach is $2.81$~\cite{Beller2013}. Another dynamic fluctuation effect, the pairing fluctuation is included explicitly in a later paper~\cite{Vaquero2013}, where an increase of about $28\%$ in the NME with respect to the previous value $1.71$ is found for ${}^{150}$Nd.

Nevertheless, our REDF results for $M^{0\nu}$ are not consistent with the NREDF calculations in Refs.~\cite{Rodriguez2010, Beller2013}. Actually, for the $0_1^+\rightarrow 0_1^+$ decay mode, the values predicted by the two EDF calculations set the upper and the lower boundaries for the calculated results. The essential difference between these two calculations is not the method, but the fact that the prolate minimum in the PEC of the nucleus ${}^{150}$Nd has a considerably smaller deformation for the relativistic functional PC-PK1 (see Fig.~\ref{fig04} of this investigation) than for the Gogny functional (see Ref.~\cite{Rodriguez2008}). This is the reason why the $E2$ transition probabilities in the spectrum of Fig.~\ref{fig06} of this paper are in much better agreement with experimental data than those obtained with the Gogny functional (see Fig. 1 of Ref.~\cite{Rodriguez2010}). In fact, the change in deformation from the initial nucleus ${}^{150}$Nd to the final nucleus ${}^{150}$Sm is considerably smaller for the functional PC-PK1 than in the Gogny case. In addition, the collective wave functions in the GCM-calculations based on the relativistic functional PC-PK1 have a considerably larger width than those obtained from the Gogny functional (see Fig.~\ref{fig05} of this paper and Fig.~1 of Ref.~\cite{Rodriguez2010}). All these lead to the fact that the transition matrix element $M^{0\nu}$ for neutrinoless double-$\beta$ decay is considerably larger in the present investigation ($M^{0\nu}=5.6$) than that obtained with the Gogny functional ($M^{0\nu}=1.7$) in Ref.~\cite{Rodriguez2010}.

Of course, so far, there is no experimental data on the value of this matrix element. Considering, however, the fact that the relativistic functional PC-PK1 reproduces the low-lying experimental spectra of ${}^{150}$Nd and ${}^{150}$Sm in a better way than the nonrelativistic functional Gogny D1S, we hope that our calculated NMEs are more reliable.  For the nucleus ${}^{150}$Nd, it is also a fact that the quantum phase transition with the X(5) character observed in the experiment of Ref.~\cite{Kruecken2002} is well reproduced by the relativistic functional PC-F1~\cite{Niksic2007}.

The half-lives $T_{1/2}^{0\nu}$ predicted by different approaches are listed in Table~\ref{tab04}, assuming the Majorana neutrino mass $\langle m_\nu\rangle=50~\mathrm{meV}$. The half-life $T^{0\nu}_{1/2}(0_1^+\rightarrow 0_1^+)$ in the present calculation turns out to be $2.1\times 10^{25}~\mathrm{y}$, which is the most optimistic prediction so far for the next generation of experiments searching for the $0\nu\beta\beta$ decay in ${}^{150}$Nd.

\section{Summary}\label{summary}

The first relativistic description for the NME of the $0\nu\beta\beta$ decay
has been given within the framework of the MR-CDFT based on a point-coupling functional PC-PK1, where the dynamic correlations related to the restoration of broken symmetries and to the fluctuations of collective coordinates are incorporated in the nuclear wave functions. For the decay candidate ${}^{150}$Nd and its daughter nucleus ${}^{150}$Sm, the low-energy spectra and electric quadrupole transitions are reproduced very well with our nuclear model. 

Comparing to other approaches, our calculations for the $0\nu\beta\beta$ decay matrix elements predict the most optimistic decay rate for ${}^{150}$Nd. Inclusion of the PNP has small
impact on the single-configuration matrix elements, while it affects the total GCM matrix element $M^{0\nu}$ with configuration mixing by changing the distributions of collective wave functions in deformation space. Consideration of the nuclear static and dynamic deformations leads to a dramatic suppression of $M^{0\nu}$ with respect to the matrix element between spherical configurations. The relativistic effects that are omitted in the nonrelativistic reduced decay operator are about $5\%$ for the ground-state to ground-state transition, and about $24\%$ for the transition from the ground state to the $0_2^+$ state. Of course, these conclusions require further systematic investigations to confirm.

\section*{Acknowledgements}
The authors thank K. Hagino, N. Hinohara, J. N. Hu, S. H. Shen, S. Q. Zhang, and P. W. Zhao and are grateful for the discussions during YITP Workshop
No. YITP-W-99-99 on ``International Molecule-type Workshop on New correlations in Exotic Nuclei and Advances of Theoretical Models.''
This work was partially supported by the Major State Basic Research Development Program of China (Grant No.~2013CB834400), the Tohoku University Focused Research Project ``Understanding the Origins for Matters in Universe,'' the National Natural Science Foundation
of China (Grants No.~11105111, No.~11175002, No.~11305134, and No.~11335002), the Research Fund for the Doctoral Program of Higher Education (Grant No.~20110001110087), the Overseas Distinguished Professor Project from Ministry of Education (Grant No.~MS2010BJDX001), the DFG cluster of excellence ``Origin and Structure of the Universe'' (www.universecluster.de), and the Fundamental Research Funds for the Central Universities (Grants No. XDJK2010B007 and No. XDJK2013C028).

\appendix

\section{Evaluation of two-body matrix elements}\label{appA}

In this section we derive explicit expressions for the TBMEs $\langle ab|\hat{O}|cd\rangle$ defined in Eqs. (\ref{operator1}) and (\ref{me}) within the closure approximation. This matrix element contains a sum over the various channels $i=VV,AA,AP,PP,MM$ and in each channel the matrix element can be expressed as an integral in momentum space over a product of single-particle matrix elements
in the following form:
\begin{eqnarray}\label{A1}
&&\langle ab|\hat{O}_i|cd\rangle=\frac{4\pi R}{g_A^2(0)}\int\frac{d^3 q}{(2\pi)^3}\frac{g_{i_1}(\bm q^2)g_{i_2}(\bm q^2)}{q(q+E_d)}~~~~~~~~~~~~~~~~\\
&&~~~~~~~~~~~~~~~~~~~~~~~~\times~\langle a|\Gamma_{i_1}\te^{\ti\bm q\bm r}| c\rangle~\langle b|\Gamma_{i_2}\te^{-\ti\bm q\bm r} |d\rangle\notag.
\end{eqnarray}
The functions $g_i(\bm q^2)$ depend on the coupling constants and the vertices $\Gamma_i$ are matrices in Dirac- and isospace given in Eq.~(\ref{twocurrentR}). For $i=P$ they also depend on the $\bm{q}$. Using $\bm q \te^{\ti\bm q\bm r}=-\ti\bm{\nabla}\te^{\ti\bm q\bm r}$ this dependence is expressed by the gradient operator.

Using the multipole expansion for plane waves~\cite{Serra2001},
\begin{eqnarray}\label{A2}
  \te^{\ti \bm q\bm r}&=& 4\pi\sum_{LM}\ti^L j_L(qr)Y^\ast_{LM}(\hat{\bm q})Y_{LM}(\hat{\bm r}),
\end{eqnarray}
and the orthonormality of spherical harmonics,
\begin{eqnarray}\label{A3}
  \int d \Omega_q Y_{LM}^*(\hat{\bm q})Y_{L'M'}(\hat{\bm q})=\delta_{LL'}\delta_{MM'},
\end{eqnarray}
we find
\begin{eqnarray}
\label{A4}
&&\langle ab|\hat{O}_{i}|cd\rangle=\frac{8 R}{g_A^2(0)}\int\frac{g_{i_1}(\bm q^2)g_{i_2}(\bm q^2){q^2\td q}}{q(q+E_d)}~~~~~~~~~~~~~~~~~~~\\
&&~~~~~\times~\sum_{LM}\langle a|\Gamma_{i_1} j_L(qr)Y_{LM}| c\rangle~\langle b|\Gamma_{i_2} j_L(qr)Y^*_{LM} |d\rangle\notag.
\end{eqnarray}

So far, the indices $a$, $b$, $c$, and $d$ characterize an arbitrary spinor basis. In a spherical basis the
single-particle spinors have the form
\begin{equation}
\label{basis}
  |1\rangle = |n_1 l_1 j_1 m_1\rangle =
  \left(
  \begin{array}{c}
  |1) \\
  \ti | \tilde{1}) \\
  \end{array}
  \right)
  \equiv
  \left(
  \begin{array}{c}
  f_{n_1}(r) |l_1j_1m_1) \\
  \ti g_{n_1}(r) |\tilde{l}_1j_1m_1) \\
  \end{array}
  \right).
\end{equation}
For clarity, here the two-dimensional spinors in spin space are expressed by round brackets.
Here the upper part $|1)$ represents the large component with the radial wave function $f_{n_1}(r)$  and the angular momentum quantum numbers $j_1l_1m_1$. The lower part $|\tilde{1})$ describes the small component with the radial wave function $g_{n_1}(r)$ and the orbital angular momentum $\tilde{l}_1=l_1 \pm 1$ for $j_1=l_1 \pm \frac{1}{2}$.

Using angular momentum coupling techniques the spin and angular parts of the matrix elements in the spherical basis can be carried out analytically.
The matrices $\Gamma_i$ contain the matrices $\gamma^0$ and $\gamma_5$ forming scalars in spin space. The products ${\gamma_\mu}^{(1)} {\gamma^\mu}^{(2)}$ are written as a scalar products of operators acting on the first and on the second particles. They have a timelike part formed by scalars and a spacelike part formed by vectors
in spin space $\bm{\gamma}=\gamma^0\bm{\alpha}=\gamma^0\gamma_5\bm{\Sigma}$ with
$\bm\Sigma=\left(
              \begin{array}{cc}
                \bm\sigma &  \\
                 & \bm\sigma \\
              \end{array}
            \right)$.
The TBMEs can be expressed in terms of scalar products of the spin operators,
\begin{equation}\label{A6}
\bm{\Sigma}^{(1)}\cdot \bm{\Sigma}^{(2)}=\sum_{M}(-)^M{\Sigma}^{(1)}_{M}{\Sigma}^{(2)}_{-M},
\end{equation}
or/and the spherical harmonics,
\begin{equation}\label{A7}
Y^{(1)}_L\cdot Y^{(2)}_L=\sum_M(-)^MY^{(1)}_{LM}Y^{(2)}_{L-M},
\end{equation}
acting on the first and on the second particles.

Recoupling the spherical operators $\bm\Sigma$ (rank 1) and $Y_{LM}$ (rank $L$) by the relation
\begin{eqnarray}
\label{A8}
&&\left(\bm\Sigma^{(1)}\cdot\bm\Sigma^{(2)}\right)\left(\,Y^{(1)}_{L}\cdot Y^{(2)}_{L}\right) \\
&&~~~~~~~~~~=~\sum_{J=L-1}^{L+1}(-)^{1+L+J} \left([\Sigma Y_L]^{(1)}_{J}\cdot [\Sigma Y_L]^{(2)}_{J}\right),\notag
\end{eqnarray}
the corresponding operators become the scalar products of single-particle operators $[\Sigma Y_L]_J$ acting on the spin and angular coordinates.

In general, the operators $\hat{O}_i$ can be expressed by scalar products of single-particle operators of rank $J$ acting on the spin and angular coordinates of the first and the second particles:
\begin{equation}\label{A9}
\hat T^{(1)}_J\cdot \hat T'^{(2)}_J=\sum_M(-)^M\hat T^{(1)}_{JM}\hat T'^{(2)}_{J-M}.
\end{equation}

Next we simplify the single-particle matrix element by using the Wigner-Eckart theorem for spherical tensor operators
of rank $J$,
\begin{equation}\label{A10}
\langle jm|\hat{T}_{JM}|j'm'\rangle=\frac{(-)^{j'-m'}}{\sqrt{2J+1}}C(jmj'-m'|JM)\langle j||T_J||j'\rangle;
\end{equation}
therefore, the angular part of TBMEs can be written as
\begin{eqnarray}\label{A11}
	&&\langle 12|\hat T^{(1)}_J\cdot \hat T'^{(2)}_J|34\rangle=\frac{1}{2J+1}(-)^{j_3-m_3}(-)^{j_4-m_2}~~~~~~~~\\
	&&~~~~~~~~~~~~~~~~~~~\times~C(j_1m_1j_3-m_3|JM)\langle 1||\hat T_J||3\rangle\notag\\
	&&~~~~~~~~~~~~~~~~~~~\times~C(j_4m_4j_2-m_2|JM)\langle 2||\hat T'_J||4\rangle.\notag
\end{eqnarray}

So far we calculated only uncoupled matrix elements. Owing to the Wigner-Eckart theorem, their $m$ dependence
is given by Clebsch-Gordan coefficients. Exploiting the orthogonality of the Clebsch-Gordan coefficients,
\begin{equation}\label{A12}
\sum_{m_1m_2}C(j_1m_1j_2m_2|JM)C(j_1m_1j_2m_2|J^\prime M^\prime)=\delta_{JJ'}\delta_{MM'}
\end{equation}
we can derive TBME coupled to good angular momentum $J$ ($ph$ coupling):
\begin{eqnarray}\label{A13}
&&\langle 12|\hat{O}|34\rangle^J_{ph} =\sum_{m_1m_3}(-)^{j_3-m_3}C(j_1m_1j_3-m_3|JM)\notag\\
&&~~~~~~~~~~~~~~~\times~\sum_{m_4m_2} (-)^{j_2-m_2}C(j_4m_4j_2-m_2|JM)\notag\\
&&~~~~~~~~~~~~~~~\times~\langle j_1m_1,j_2 m_2|\hat{O}|j_3 m_3,j_4 m_4\rangle.~~~~~~~~~~~~~~~
\end{eqnarray}
We finally obtain for the spin and angular part of the different TBMEs
\begin{equation}
\label{A14}
\langle 12 |\hat{T}_J^{(1)}\cdot\hat T'^{(2)}_J|34\rangle^J_{ph}=\frac{(-)^{j_4-j_2}}{2J+1}\langle 1 ||\hat{T}_J|| 3 \rangle \langle 2 ||\hat T'_J|| 4 \rangle.
\end{equation}

The reduced matrix elements for the operators $Y_L$ and $[\sigma Y_L]_J$ are given by
\begin{eqnarray}\label{A15}
  &&( l_1j_1||Y_L||l_2j_2) =(-)^{j_1-j_2}( l_2j_2||Y_L||l_1j_1) \\
  &=&\frac{1+(-)^{l_1+l_2+L}}{2}\frac{\hat j_1\hat j_2\hat L}{\sqrt{4\pi}}
   (-)^{L+j_2-\frac12}\left(
                      \begin{array}{ccc}
                        j_1 & L & j_2 \\
                        -\frac12 & 0 & \frac12 \\
                      \end{array}
                    \right),\notag
\end{eqnarray}
and
\begin{eqnarray}\label{A16}
&&(l_1j_1||[\sigma Y_L]_J||l_2j_2)=(-)^{j_1+j_2+L+J}( l_2j_2||[\sigma Y_L]_J||l_1j_1)\notag\\
&=& \frac{1+(-)^{l_1+l_2+L}}{2}\frac{\hat j_1\hat j_2\hat L\hat J}{\sqrt{4\pi}} (-)^{l_2+j_1+j_2+L+1}\notag\\
&\times&\left[(-)^{l_2+j_2+\frac12}\left(
                                           \begin{array}{ccc}
                                             1 & L & J \\
                                             0 & 0 & 0 \\
                                           \end{array}
                                         \right)
                                         \left(
                                           \begin{array}{ccc}
                                             j_1 & L & j_2 \\
                                             \frac12 & 0 & -\frac12 \\
                                           \end{array}
                                         \right)\right.\notag\\
   &&~~~-~\left.\sqrt{2}\left(
                 \begin{array}{ccc}
                   1 & L & J \\
                   -1 & 0 & 1 \\
                 \end{array}
               \right)\left(
                 \begin{array}{ccc}
                   j_1 & J & j_2 \\
                   \frac12 & -1 & \frac12 \\
                 \end{array}
               \right)\right].
    \end{eqnarray}
Here $\hat j=\sqrt{2j+1}$. Note that an extra phase factor $(-)^{(l_1+1/2-j_1)+(l_2+1/2-j_2)}$ is added to the reduced matrix elements given in Ref.~\cite{Serra2001}, because orbit-spin ($LS$) coupling instead of spin-orbit ($SL$) coupling for the single-particle states is used throughout the calculation.

For the radial part, the radial integrals $(nl| j_L(qr)|n'l')$ for spherical oscillator wave functions
are treated in Sec.~6 of this appendix. Of course, in Eq.  (\ref{sum}) we need the $pp$-coupled matrix elements.
They are obtained from the $ph$-coupled matrix elements by recoupling~\cite{Serra2001}
\begin{eqnarray}\label{A17}
&&\langle 12|\hat{O}|34\rangle^{\lambda}_{pp}=\sum_J (2J+1)(-)^{j_3+j_4+\lambda}~~~~~~~~~~~~~~~~\\
&&~~~~~~~~~~~~~\times~\left\{
                                    \begin{array}{ccc}
                                      j_1 & j_2 & \lambda \\
                                      j_4 & j_3 & J \\
                                    \end{array}
                                  \right\}
\langle 12|\hat{O}|34\rangle^{J}_{ph}.\notag
\end{eqnarray}
In the end, we return to the uncoupled matrix elements by
\begin{eqnarray}\label{A18}
    &&\langle 12|\hat O|34\rangle=\sum_{\lambda (M)}C(j_1m_1j_2m_2|\lambda M)~~~~~~~~~~~~~~~~~~~~\\
    &&~~~~~~~~~~~~~~~\times~C(j_3m_3j_4m_4|\lambda M)\langle 12|\hat O|34\rangle^{\lambda}_{pp}.\notag
\end{eqnarray}

In detail we obtain the following $ph$-coupled matrix elements (\ref{A13}) for the different channels of Eq.~(\ref{twocurrentR}). For the sake of
simplicity, in the following coupled matrix elements a common factor  $8 R/(g_A^2(2J+1))$, as well as a common phase $(-)^{j_4-j_2}$, are left out.

\subsection{Vector coupling term $\hat O_\mathrm{VV}$}
For VV we have  in Eq. (\ref{VV}) the vertex $\Gamma_V = \gamma^0\gamma_\mu$ (neglecting the isospin operator) and therefore, using
Eq. (\ref{A14}) we obtain the $ph$-coupled TBME,
\begin{eqnarray}\label{NMEVV}
&&\langle 12|\hat{O}_{VV}|34\rangle^J_{ph}=\int\frac{g_V^2(\bm q^2){q^2\td q}}{q(q+E_d)}~~~~~~~~~~~~~\\
&&~~~~~~~~~\times~\left( A^J_{13} A^J_{24} - \sum_L (-)^{1+L+J}  B^{L,J}_{13} B^{L,J}_{24}\right),\notag
\end{eqnarray}
with the integrals
\begin{eqnarray}
\label{AJ}
A^J_{13}~&=& \langle 1|| j_J(qr)  Y_J || 3\rangle\\
&=&(1|j_J|3) ( 1|| Y_J ||3)+(\tilde{1}|j_J|\tilde{3})(\tilde{1}|| Y_J||\tilde{3}),\notag \\
\label{BLJ}
B^{L,J}_{13} &=&\langle 1| j_L(qr) \gamma_5[\Sigma Y_L]_J  || 3\rangle\\
&=&
\ti(1|j_L|\tilde{3} )(1 ||[\sigma Y_L]_J|| \tilde{3})
-\ti(\tilde{1}|j_L|3)(\tilde{1} ||[\sigma Y_L]_J||3),\notag
\end{eqnarray}
with the reduced matrix elements given in Eqs.~(\ref{A15}) and (\ref{A16}). Note that the phase $(-)^{1+L+J}$ appearing before $B^{L,J}_{13} B^{L,J}_{24}$ comes from the recoupling of the spherical operators in Eq.~(\ref{A8}).

\subsection{Axial-vector coupling term $\hat O_\mathrm{AA}$}
For AA coupling we have  in Eq. (\ref{AA}) the vertex $\Gamma_A = \gamma^0\gamma_\mu\gamma_5$ and, therefore, using
Eq. (\ref{A14}) we obtain the $ph$-coupled TBME
\begin{eqnarray}\label{NMEAA}
&&\langle 12|\hat{O}_{AA}|34\rangle^J_{ph}=\int\frac{g_A^2(\bm q^2){q^2\td q}}{q(q+E_d)}~~~~~~~~~~~~~\\
&&~~~~~~~~~\times~\left( C^J_{13} C^J_{24} - \sum_L (-)^{1+L+J}D^{L,J}_{13} D^{L,J}_{24}\right),\notag
\end{eqnarray}
with the integrals
\begin{eqnarray}
\label{CJ}
C^J_{13}~&=&  \langle 1||j_J(qr)\gamma_5 Y_J || 3\rangle\\
&=&\ti( 1| j_J|\tilde{3})( 1 || Y_J ||\tilde{3})
- \ti( \tilde{1}|j_J|3)( \tilde{1} || Y_J||3),\notag \\
\label{DLJ}
D^{L,J}_{13} &=&\langle 1|| j_J(qr)[\Sigma Y_L]_J  || 3\rangle\\
&=&( 1| j_L|3 ) (  1 ||[\sigma Y_L]_J||3)
+( \tilde{1}| j_L|\tilde{3} ) ( \tilde{1} ||[\sigma Y_L]_J||\tilde{3}).\notag
\end{eqnarray}

\subsection{Axial-vector and pseudoscalar coupling term $\hat O_\mathrm{AP}$}
For the TBME of the AP coupling term $\langle 12| \hat O_{AP}|34\rangle$ we have
in the $q$ integral the matrix elements [Eq.~(\ref{AP})]:
\begin{equation}\label{NMEAP1}
\langle 1|\gamma^0\bm{\gamma}\gamma_5\te^{\ti\bm q\bm r}| 3\rangle\cdot\langle 2|\gamma^0\gamma_5 \bm q\te^{-\ti\bm q\bm r} |4\rangle.
\end{equation}
Because $\bm q \te^{\ti\bm q\bm r}=-\ti\bm{\nabla}\te^{\ti\bm q\bm r}$, we obtain
\begin{equation}\label{NMEAP2}
\sum_{J}-\ti\langle 1|(\bm\Sigma\cdot\bm{\nabla})j_J(qr)Y_{J}| 3\rangle \cdot\langle 2|\gamma^0\gamma_5 j_J(qr)Y_{J}|4\rangle.
\end{equation}
It can be proved that
\begin{eqnarray}\label{relation1}
   &&\bm \Sigma\cdot \bm{\nabla} j_J(qr)Y_{JM}=\sqrt{\frac{J+1}{2J+1}}qj_{J+1}(qr)[\Sigma Y_{J+1}]_{JM}\notag\\
   &&~~~~~~~~~~~~+\sqrt{\frac{J}{2J+1}}qj_{J-1}(qr)[\Sigma Y_{J-1}]_{JM}.
\end{eqnarray}
Therefore, in a spherical basis we find for the coupled matrix element
\begin{eqnarray}\label{NMEAP}
&&\langle 12|\hat{O}_{AP}|34\rangle^J_{ph} = 2\int\frac{g_A(\bm{q}^2)g_P(\bm q^2)q^3\td q}{q(q+E_d)}\\
&&\times~(-\ti)\left( \sqrt{\frac{J+1}{2J+1}}D_{13}^{J+1,J} +\sqrt{\frac{J}{2J+1}}D^{J-1,J}_{13} \right)E^J_{24},\notag
\end{eqnarray}
with the integral $D^{L,J}_{13}$ in Eq. (\ref{DLJ}) and the integral
\begin{eqnarray}
\label{EJ}
E^J_{13}~&=& \langle 1|| j_J(qr)  \gamma^0\gamma_5 Y_J || 3\rangle\\
&=&\ti(1|j_J|\tilde{3}) ( 1 || Y_J ||\tilde 3)
+\ti(\tilde{1}|j_J|3)(\tilde{1} || Y_J||3).\notag
\end{eqnarray}

\subsection{Pseudoscalar coupling term $\hat O_\mathrm{PP}$}
For PP coupling we have in Eq. (\ref{PP}) the vertex $\Gamma_P = \bm{q}\gamma^0\gamma_5$ and, therefore, using
Eq. (\ref{A14}) we obtain the $ph$-coupled TBME,
\begin{equation}\label{NMEPP}
\langle 12|\hat{O}_{PP}|34\rangle^J_{ph}=\int\frac{g_P^2(\bm q^2){q^4\td q}}{q(q+E_d)}E^J_{13} E^J_{24},
\end{equation}
with the integral $E^J_{13}$ given in Eq.~(\ref{EJ}).

\subsection{Weak-magnetism coupling term $\hat O_\mathrm{MM}$}
For the TBME of the MM coupling term $\langle 12| \hat O_{MM}|34\rangle$ we have
in the $q$-integral the matrix elements [Eq.~(\ref{MM})]
\begin{equation}\label{NMEMM1}
\langle 1|\gamma^0\sigma_{\mu i}q^i\te^{\ti\bm q\bm r}| 3\rangle~\langle 2|\gamma^0\sigma^{\mu j}q_j\te^{-\ti\bm q\bm r} |k\rangle.
\end{equation}
Using the definition of the Dirac matrix
\begin{eqnarray*}
  \sigma_{\mu\nu}=\frac{\ti}{2}\left[\gamma_\mu,\gamma_\nu\right]\quad\text{or}
  \quad\sigma_{0i}=\ti\alpha_i,\quad
  \sigma_{ij}=\varepsilon_{ijk}\Sigma^k,
\end{eqnarray*}
we have
\begin{eqnarray*}
  \sigma_{0i}q^i=\ti\bm\alpha\cdot\bm q,\quad \sigma_{ki}q^i=-\left[\bm\Sigma\times\bm q\right]_k.
\end{eqnarray*}
Making use of
\begin{eqnarray}\label{relation2}
&&\left(\bm\Sigma^{(1)}\times\bm q\right) \left(\bm\Sigma^{(2)}\times\bm q\right)\\
&&~~~~~~=~\left(\bm\Sigma^{(1)}\cdot\bm\Sigma^{(2)}\right)q^2-\left(\bm\Sigma^{(1)}\cdot\bm q\right) \left(\bm\Sigma^{(2)}\cdot\bm q\right),\notag
\end{eqnarray}
and replacing $\bm{q}$ by the gradient we find three terms:
\begin{itemize}
\item[(1)] $\ti\left(\bm\alpha\cdot\bm q\right)$ leads to the vertex $\gamma^0\gamma_5\left(\bm{\Sigma}\cdot\bm{\nabla}\right)$;
\item[(2)]  $q^2 \left(\bm\Sigma^{(1)}\cdot\bm\Sigma^{(2)}\right)$ is to be recoupled and leads to the vertex
$q\gamma^0[\Sigma Y_L]_{J}$ [for details, see Eq. (\ref{A8})];
\item[(3)] a term with the vertex $\gamma^0\left(\bm\Sigma\cdot\bm{\nabla} \right)$.
\end{itemize}
Therefore, in a spherical basis we find for the coupled matrix element
\begin{eqnarray}\label{NMEMM}
&&\langle 12|\hat{O}_{MM}|34\rangle^J_{ph} =\frac{1}{4m^2_p}\int\frac{g_M^2(\bm q^2)q^4\td q}{q(q+E_d)}\\
&&~\left\{~\ti^2\left( \sqrt{\frac{J+1}{2J+1}}F_{13}^{J+1,J} +\sqrt{\frac{J}{2J+1}}F^{J-1,J}_{13} \right)\right.\notag\\
&&~~~~\times\left( \sqrt{\frac{J+1}{2J+1}}F_{24}^{J+1,J} +\sqrt{\frac{J}{2J+1}}F^{J-1,J}_{24} \right)\notag\\
&&~-~~~\sum_L (-)^{(1+L+J)}G^{L,J}_{13} G^{L,J}_{24}\notag\\
&&~+~\left( \sqrt{\frac{J+1}{2J+1}}G_{13}^{J+1,J} +\sqrt{\frac{J}{2J+1}}G^{J-1,J}_{13} \right)\notag\\
&&~~~\left.\times\left( \sqrt{\frac{J+1}{2J+1}}G_{24}^{J+1,J} +\sqrt{\frac{J}{2J+1}}G^{J-1,J}_{24} \right)\right\}\notag
\end{eqnarray}
with the integrals
\begin{eqnarray}
\label{FLJ}
F^{L,J}_{13}&=&\langle 1||j_L(qr)\gamma^0\gamma_5[\Sigma Y_L]_J||3\rangle\\
&=&\ti({1}|j_{L}|\tilde 3)({1}||[\sigma Y_{L}]_J||\tilde 3)
+ \ti(\tilde 1| j_{L}|{3}) (\tilde 1||[\sigma Y_{L}]_J|| 3),\notag\\
G^{L,J}_{13}&=&\langle 1||j_L(qr)\gamma^0[\Sigma Y_L]_J||3\rangle\\
&=&({1}|j_{L}| 3)({1}|[\sigma Y_{L}]_J|| 3)
- (\tilde 1| j_{L}|\tilde{3}) (\tilde 1||[\sigma Y_{L}]_J||\tilde 3).\notag
\end{eqnarray}

\bigskip

\subsection{Slater integrals}
\label{Slater}

From previous appendices, we have seen that the Slater integrals in the TBMEs read
\begin{eqnarray}
	{S^{L_1L_2}_{1234}\equiv\int \td q D(q)\langle 1|j_{L_1}(qr)|3\rangle\langle 2|j_{L_2}(qr)|4\rangle.}
\end{eqnarray}
Here $|k\rangle$ represent an arbitrary set radial wave functions (for the large or small components). In the SHO basis these
integrals can be evaluated analytically (see Ref.~\cite{Serra2001}),
\begin{eqnarray}
  &&S^{L_1L_2}_{n_1l_1n_2l_2n_3l_3n_4l_4} \\
  &=&\int \td q D(q)\langle n_1l_1|j_{L_1}(qr)|n_3l_3\rangle\langle n_2l_2|j_{L_2}(qr)|n_4l_4\rangle\notag\\
  &=& \frac{\pi}{8}\sum_{N_1=N_{m1}}^{N_{M1}}\sum_{N_2=N_{m2}}^{N_{M2}}{A_{n_1l_1n_3l_3}^{N_1 L_1}A_{n_2l_2n_4l_4}^{N_2L_2}}\notag\\
  &\times&{b^{3}} \int \td q D(q)\te^{-b^2q^2/4}R_{N_1L_1}({\frac{b^2q}{2}})
  R_{N_2L_2}({\frac{b^2q}{2}}),\notag
\end{eqnarray}
where $N_{m1}={(l_1+l_3-L_1)}/{2}$ and $N_{M1}=n_1+n_3+N_{m1}$. $R_{nl}(r/b)=\langle r|nl\rangle$ represent spherical radial oscillator wave functions, $b$ is the oscillator length, $D(q)$ indicates a function of $q$, and the coefficients $A_{nln'l'}^{NL}$ are given by
\begin{eqnarray}
&&A_{nln'l'}^{NL}=\sqrt{n!(n+l+\frac12)!}\sqrt{n'!(n'+l'+\frac12)!}\notag\\
&\times&\sqrt{N!(N+L+\frac12)!}\sum_{q,q'=0}^{n,n'}\notag\\
&\times&\frac{\delta_{0,q+q'-N+N_m}{(-)^{N-N_m}}}{q!q'!(n-q)!(n'-q')!(q+l+\frac12)!(q'+l'+\frac12)!}.\notag\\
\end{eqnarray}


\end{document}